\documentclass{aa}
\usepackage{graphicx}

\begin{document}

\newcommand{\hi}{\ion{H}{i}~}
\newcommand{\hii}{\ion{H}{ii}~}

   \title{\hi shells in the outer Milky Way} 

   \author{S. Ehlerov\' a 
           \and
           J. Palou\v s 
           }

   \offprints{S. Ehlerov\'a }

   \institute{Astronomical Institute,
         Academy of Sciences of the Czech Republic,
         Bo\v cn\' \i \ II 1401, 141 31 Prague 4, Czech Republic \\
              }

   \date{Received 24 September 2003/ Accepted 7 March 2005}

   \titlerunning{\hi shells in the outer Milky Way}
   \authorrunning{Ehlerov\'a \& Palou\v s}

\abstract{
We present results of a method for an automatic search for \hi shells 
in 3D data cubes and apply it to the Leiden-Dwingeloo \hi survey 
of the northern Milky Way. In the 2nd Galactic quadrant, where 
identifications of structures are not substantially influenced by 
overlapping, we find nearly 300 structures. 
The Galactic distribution of shells has an exponential profile in the 
radial direction with a scale length of $\sigma_{\mathrm{gsh}} = 3$ kpc. 
In the $z$ direction, one half of the shells are found at distances 
smaller than 500 pc. We also calculate the energies necessary to create 
the shells: there are several structures with energies greater than 
$10E_{\mathrm{SN}}$ but only one with an energy exceeding 
$100E_{\mathrm{SN}}$. Their size distribution, corrected for distance
effects, is approximated by a power-law with an index $\alpha = 2.1$. 
Our identifications provide a lower limit to the filling factor of shells 
in the outer Milky Way: $f_{\mathrm{2D}} = 0.4$ and $f_{\mathrm{3D}} = 0.05$.
}
\maketitle


\section{Introduction} 

The interstellar medium (ISM) in galaxies is far from homogeneous: 
it is turbulent and composed of several coexisting phases. This 
creates a complex picture, so far not fully understood. Important 
constituents of this picture are structures known as shells, supershells
and holes (usually called this in association with observations in \hi) 
or bubbles and superbubbles (in connection with $H_{\alpha}$ or other 
wavelengths). These structures are known to exist in the Milky Way Galaxy 
and many external galaxies (for a summary of observations of shells see 
Walter \& Brinks, 1999). Most probably they are created by energy 
release from massive stars (winds and supernova explosions), however, 
alternative explanation, gamma-ray bursts (GRBs; Efremov et al., 1998; 
Loeb \& Perma, 1998) or high velocity cloud (HVC) infalls (Tenorio-Tagle 
\& Bodenheimer, 1988), have been invoked in some cases. The majority of the 
observed shells are due to star formation.  Structures connected with the 
infall of HVC are exceptional (Ehlerov\' a \& Palou\v s, 1996). The origin 
of shells in connection with GRBs has not been proven (Efremov et al., 1999). 
A connection between massive stars and \hi shells can be tested either by 
observation of individual objects or by a comparison of statistical properties
such as galactic or size distributions. In this paper we choose the 
second method.
  
There are many papers on shells in the Milky Way but the majority 
of them are concerned with individual objects. The most notable exceptions are 
two papers by Heiles (1979, 1984) which contain lists of shells in the 
northern Galaxy, the paper by Hu (1981) on northern high latitude shells, 
and more recently the paper by McClure-Griffiths et al. (2002) on shells 
in the southern Galaxy. Unfortunately, none of these lists are complete, but 
are the best data available for the Milky Way. In this paper we try to create 
a more complete list of shells in our Galaxy. We use the Leiden-Dwingeloo 
\hi survey and create an automatic algorithm to find shells. Then we compare 
identified structures with shells in external galaxies and discuss their 
Galactic and size distributions.

The structure of the paper is as follows: we start with a short description 
of the dataset used for identification (Sect. 2), continue with a 
description of the identification method (Sect. 3) and other searching 
methods (Sect. 4). Then we analyse results for the 2nd galactic quadrant 
of the  Milky Way (Sects. 5 and 6) and summarise our findings 
(Sect. 7). In the Appendix we try to cross-identify the shells found 
with those published previously.

\section{Data}

We use the Leiden-Dwingeloo \hi survey (Hartmann \& Burton 1997; LDS).
The survey was made in 1989-1993 with a 25m radio telescope in 
Dwingeloo (the Netherlands). It covers 79\% of the sky with a spatial 
resolution of $0.5^{\circ}$, a velocity resolution of 
$1.03\ \mathrm{kms^{-1}}$ and an rms noise of $0.07\ {\mathrm{K}}$.

We chose this survey because of high sky coverage and uniformity 
of data. LDS has not been searched for shells up to now. Contrary to many 
\hi surveys used to find shells, like the Canadian Galactic Plane 
Survey (CGPS), LDS covers the whole sky, not only a strip around the 
Galactic equator. Therefore we are able to study shells with large angular 
sizes ($\Delta b > 10^{\circ}$) and also the distribution of shells at 
high latitudes. 

The automatic method of shell detection that we describe in this 
paper is quite general and can be applied to any data cube. It is obvious 
that by using a data cube with a higher angular resolution (such as CGPS) we 
would discover more smaller structures than by using LDS. However, as it 
will be shown later (e.g. Fig. \ref{figpose}), angular dimensions of many 
discovered shells exceed the region covered by CGPS. Consequently, they would 
remain undiscovered if we did not use an all-sky survey. Moreover, a structure 
with a dimension of 500 pc (which is quite typical) has an angular dimension 
larger than $5^{\circ}$ at a heliocentric distance less than 5.7 kpc. 
Therefore, as we are primarily interested in shells with dimensions of 
several 100 pc, the LDS (or any other high-coverage, low-resolution survey) is 
appropriate. If we were interested in smaller shells, like wind-blown
bubbles, a high-resolution survey (such as CGPS) would be better.

Our position inside the Milky Way makes the identification more difficult 
compared to external galaxies because of substantial overlapping of material 
along the line-of-sight, because of a linear resolution that changes with
heliocentric distance and because of problems with distance determination.

\section{Searching algorithm} 

\begin{figure}
 \centering
 \includegraphics[angle=270,width=8.0cm]{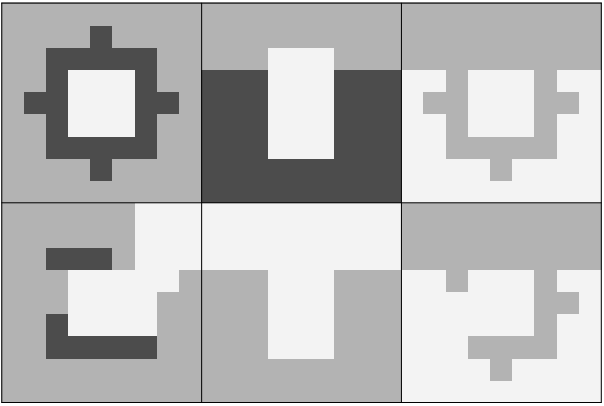}
 \caption{Types of \hi shells. The first row shows examples
          of structures, that can be identified, the second row shows 
          structures connected to the surroundings or with 
incomplete walls, which are undetectable by the code.}
 \label{figcrt1}
\end{figure}

A prototypical ideal \hi shell has several signatures which
distinguish it from the background: it is a region of decreased \hi emission 
surrounded by a dense thin wall, it is expanding and its shape is spherical. 
The reality is more complex: the ISM is stratified and turbulent; \hi shells 
evolving in this medium are elongated and irregular, their expansion slows 
down during the evolution; and there are usually intervening material, 
gradients induced by galactic rotation, and the finite resolution of 
the telescope, all of which render the identification of shells more 
difficult. Taking the difficulties mentioned above into consideration 
we conclude that the most persistent quality of an \hi shell is the 
existence of a region with decreased \hi emission, e.g. an \hi hole. Our 
automatic algorithm is based on this conclusion and it searches for holes.

The identification procedure has three steps: 1) a search in separate
velocity channels, 2) a merging of structures in adjacent velocity
channels to form an \hi shell candidate, and 3) a check using the
central pixel spectrum.

\begin{figure*}
 \centering
 \includegraphics[angle=0,width=5.5cm]{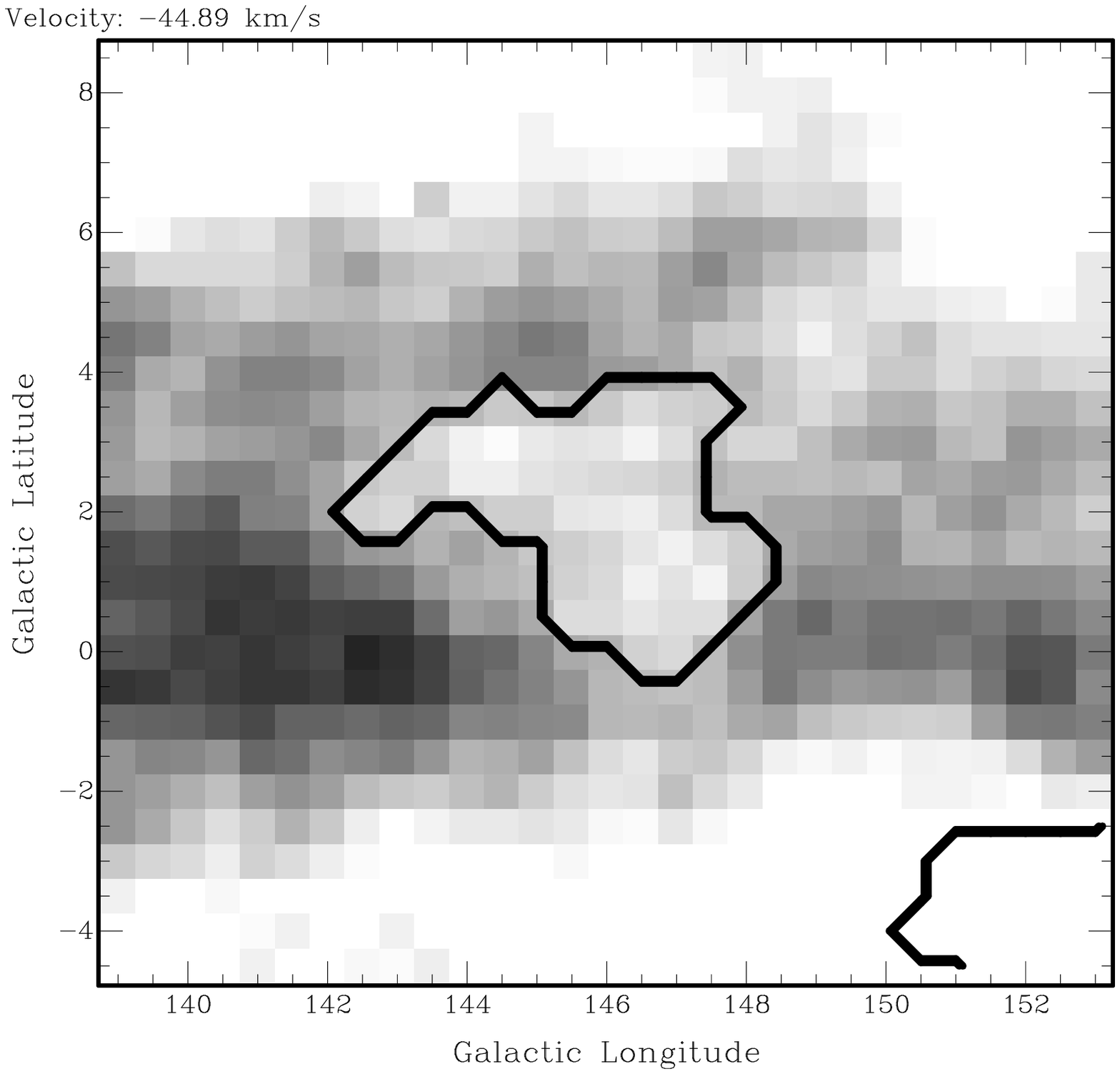}
 \includegraphics[angle=0,width=5.5cm]{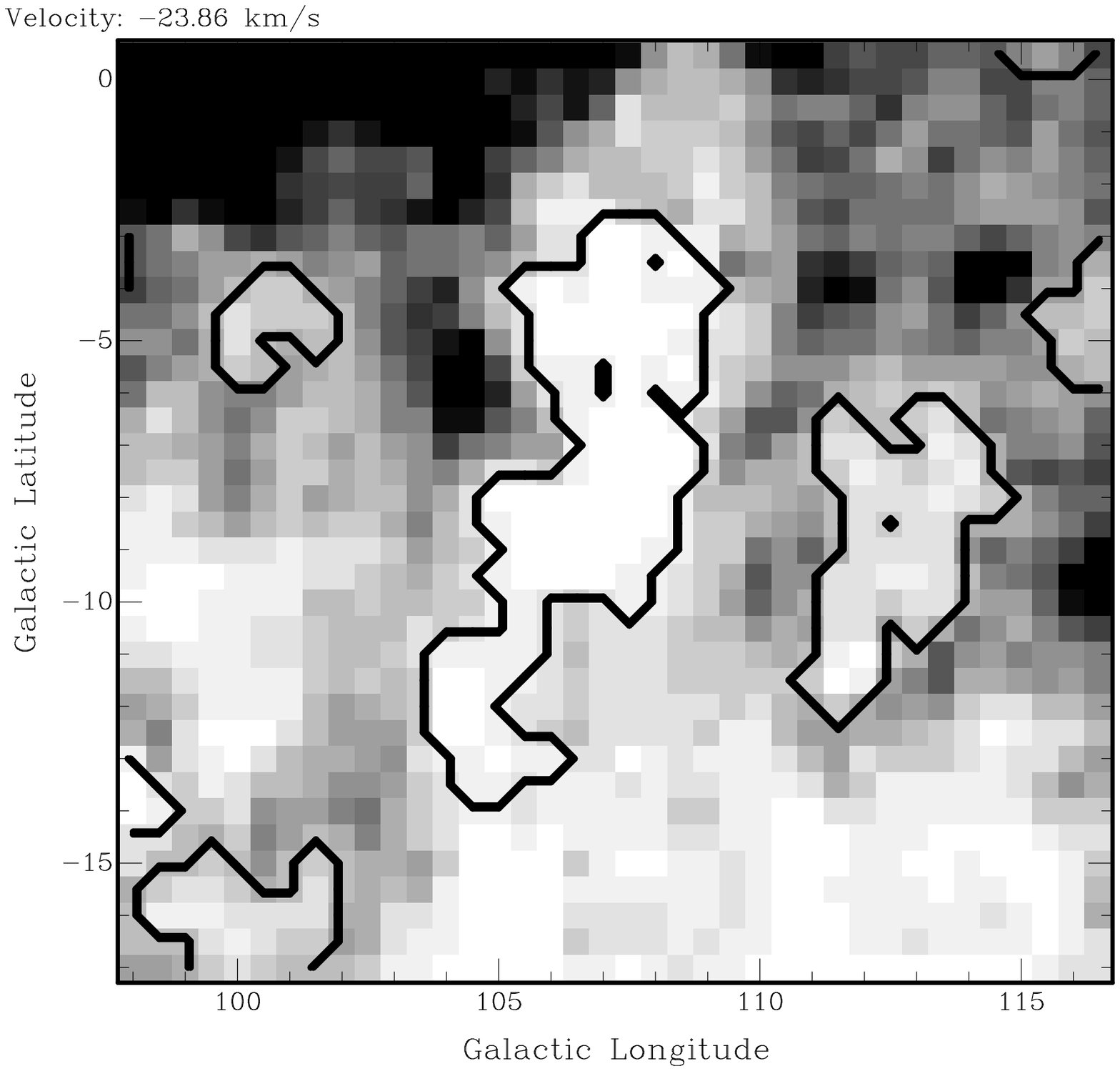}
 \includegraphics[angle=0,width=5.5cm]{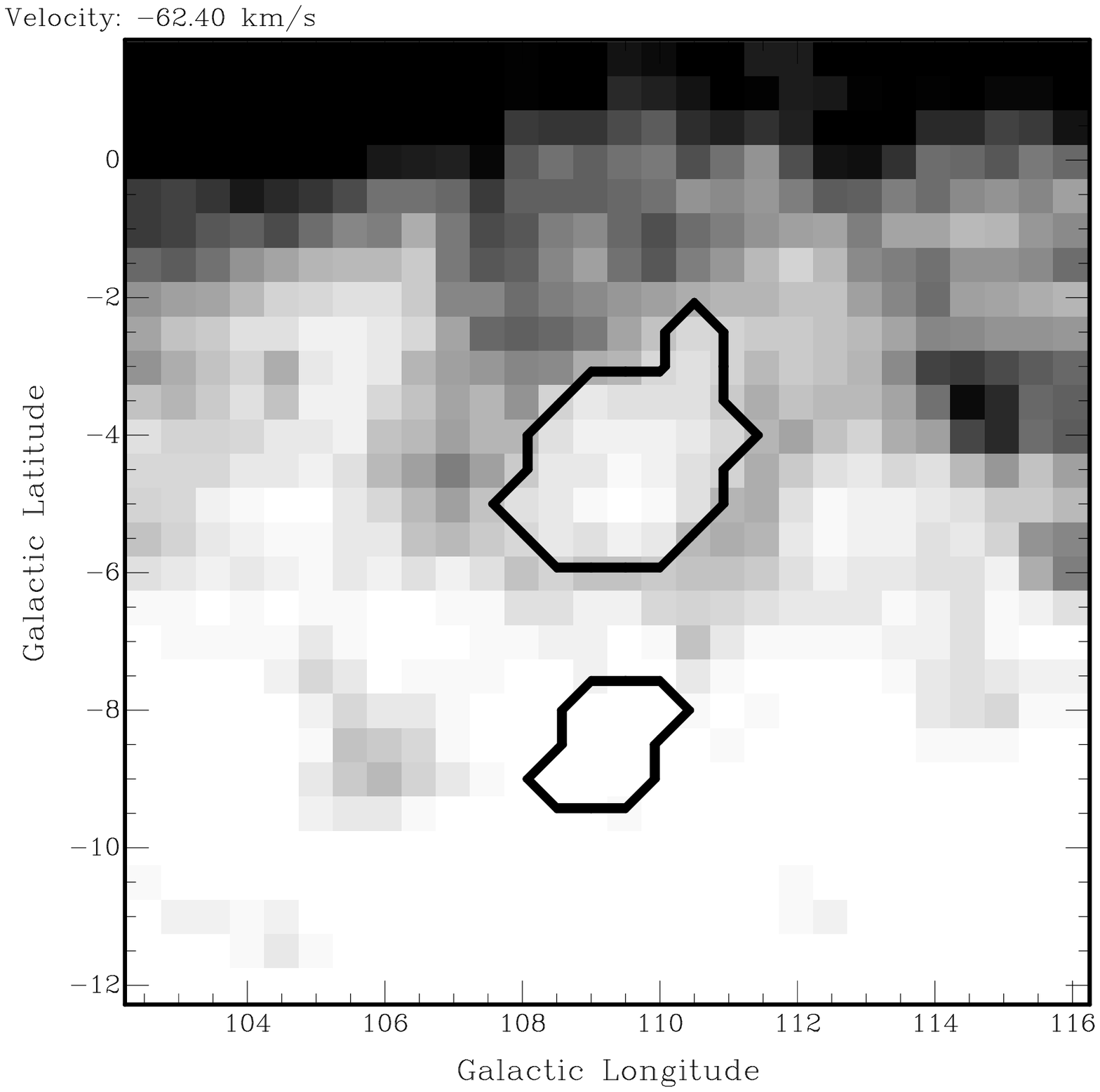}
 \includegraphics[angle=0,width=5.5cm]{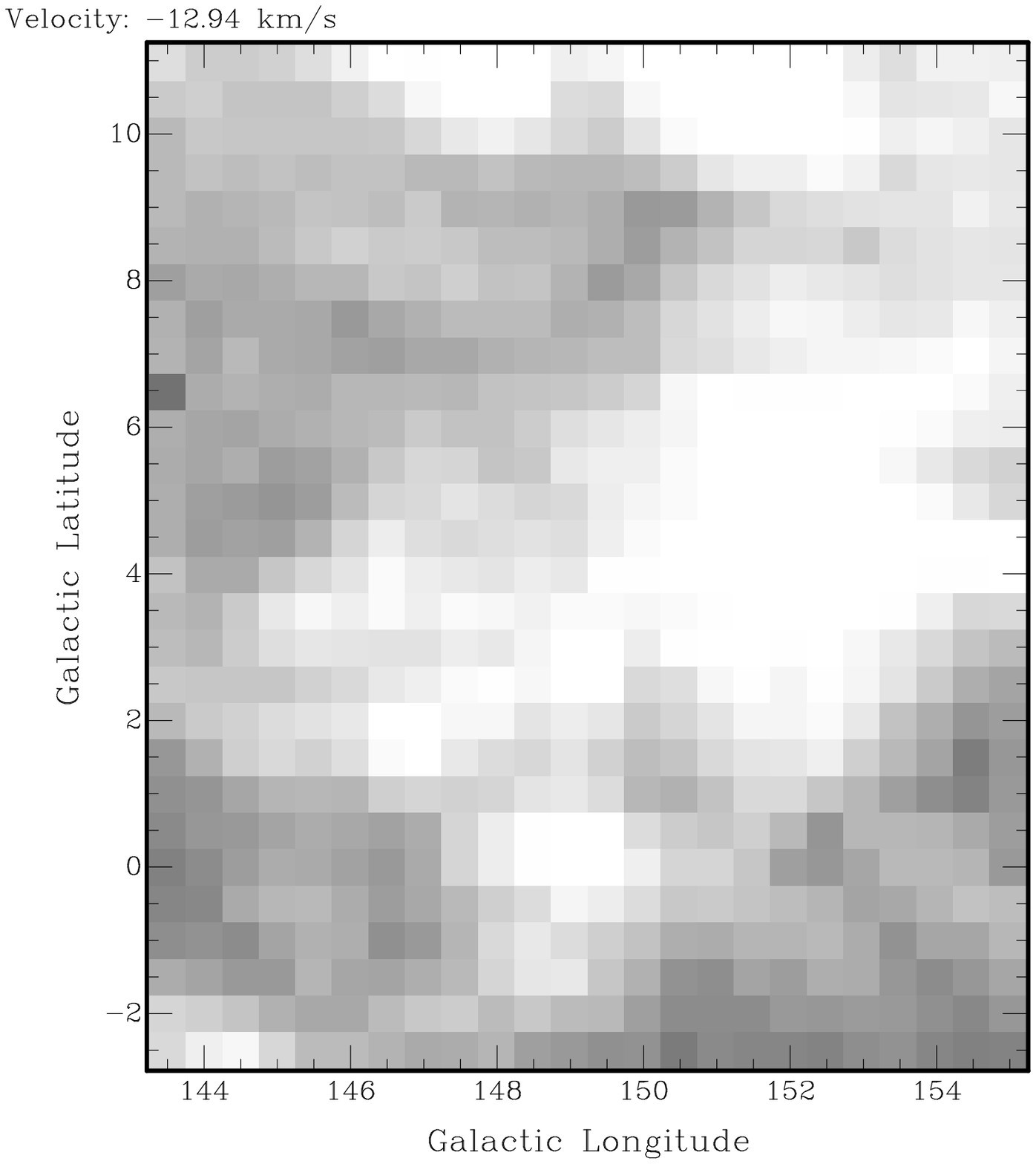}
 \includegraphics[angle=0,width=5.5cm]{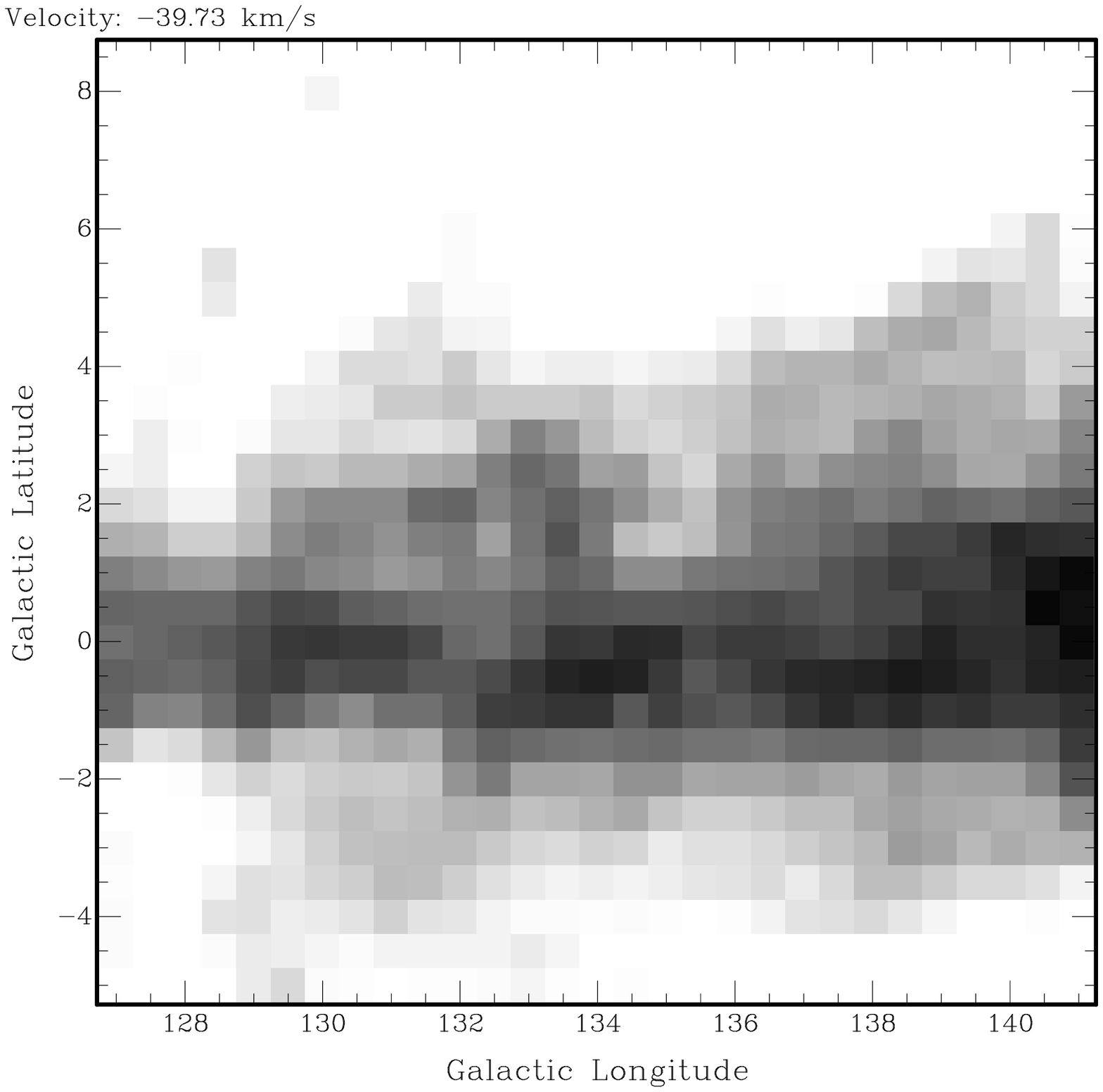}
 \includegraphics[angle=0,width=5.5cm]{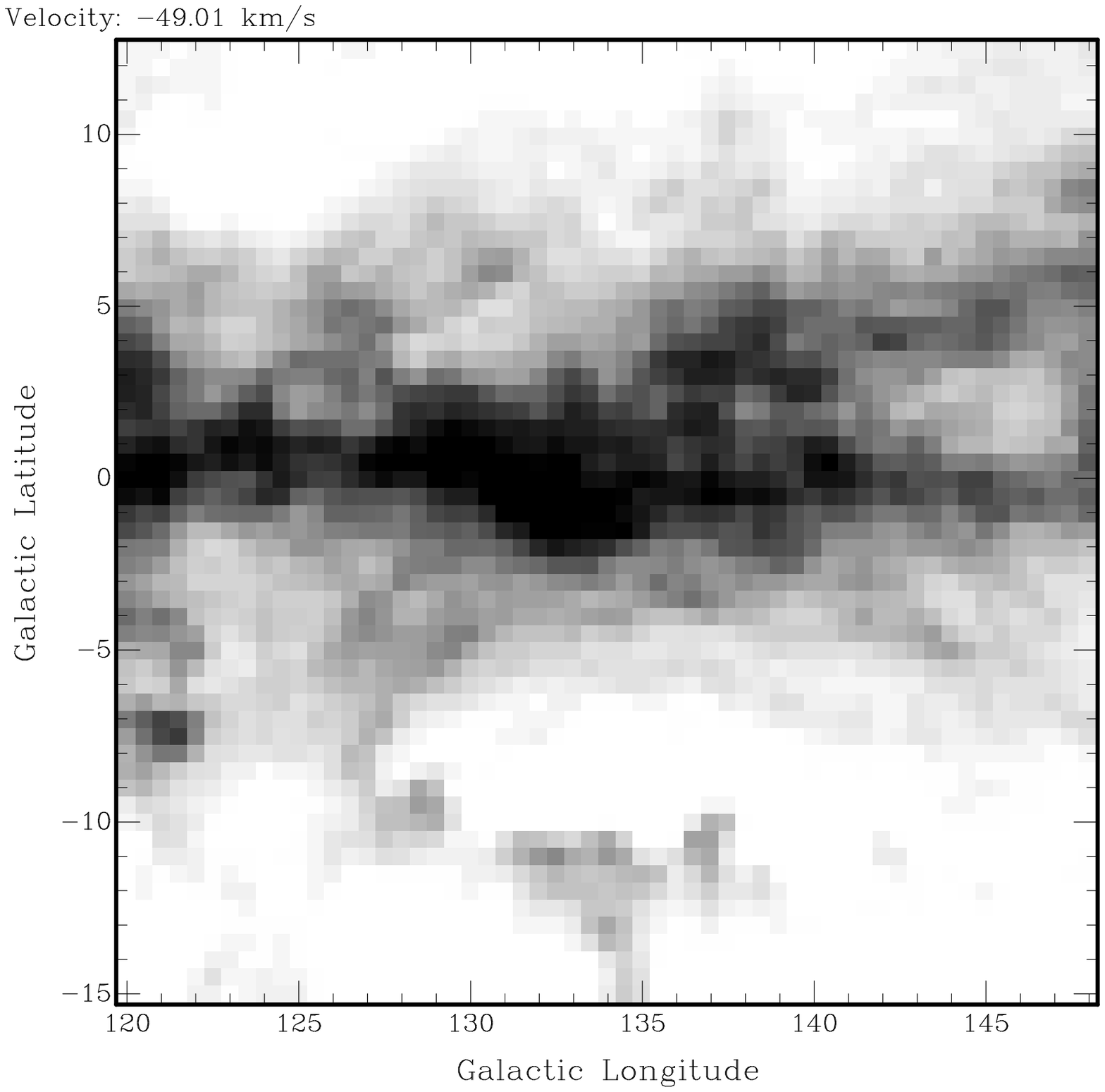}
 \caption{\hi shells in the LDS corresponding to different types as
          shown in Fig. \ref{figcrt1}.}
 \label{figtyplds}
\end{figure*}

\subsection{The search for holes in separate velocity channels} 

We search for regions of local low brightness temperature completely 
encircled by higher temperature surroundings. Not all pixels inside the 
encircled region must belong to the hole, but all participating pixels must
form a continuous structure. For each local minimum we search for the maximum 
encircling temperature, i.e. the one defining the maximum region that
fulfills the described condition. Fig. \ref{figcrt1} shows different types 
of prototypical structures; in the upper row there are structures 
detectable with our algorithm, in the lower row there are structures that
are not detectable. The first case is a local depression surrounded by a
wall, which has a higher density than the medium around it 
(Fig. \ref{figcrt1}, upper left). In some cases the wall is not observed 
and the structure is just a local depression in the \hi gas. The second 
case (Fig. \ref{figcrt1}, upper center) is a local depression in the medium 
with a steep density gradient. This case is known as a galactic \hi worm and 
the structure is frequently perpendicular to the galactic disc because of  
the large--scale $z$-stratification of the \hi. The last type is an arc blown 
out into a low density environment (Fig. \ref{figcrt1}, upper right). 

All three described  cases are detectable with our search algorithm. The 
difficulty begins when the local depression is directly connected to other 
low density regions (Fig. \ref{figcrt1}, lower left). The search 
algorithm is unable to close the structure. Thus, it either joins it to a 
larger
neighboring hole or does not identify it at all. Worms, which are open 
at one or both ends connecting regions where the \hi emission is absent 
(Fig. \ref{figcrt1}, lower center) are undetected as well. The identification
also fails for fragmented arcs in a medium with low \hi emission 
(Fig. \ref{figcrt1}, lower right). Examples from the LDS of the cases
described are given in Fig. \ref{figtyplds}.  

The only constraints put on the hole are its dimensions: the hole must be 
greater than a minimum dimension and smaller than a maximum dimension. We use 
the range of dimensions ($1.5^{\circ}$, $45^{\circ}$). There are no 
constraints on the shape of \hi  holes.

In this paper we search in $lb$ channel maps. $PV$ diagrams can be 
used as well, however, many shells are open in velocity diagrams (i.e. only 
one hemisphere is visible) and this kind of structure is not well suited 
to our type of search. A full 3D search is time-consuming and also suffers 
from the mentioned disadvantages of $PV$ diagrams.

\subsection{The formation of an \hi shell candidate} 

After finding holes, i.e. continuous regions around local minima in individual
velocity channels, substantially overlapping holes in consecutive radial 
velocity channels are automatically joined together to form a 3D 
structure. If the structure joins four or more velocity channels (i.e. if 
its velocity extent is larger than 4 kms$^{-1}$), it is called an \hi shell 
candidate. In the majority of cases the shell extends also to a few 
adjacent velocity channels, in which the angular size is less than 
1.5$^\circ $, and therefore it is not indentified by our algorithm in 
step 1 (but it can be identified in step 3, see below). If only four 
or five consecutive velocity channels are involved, the velocity 
difference between parts of the HI shell candidate is less than the typical 
velocity dispersion in the ISM: $\sigma _{\mathrm{ISM}}$ = 5 - 7 km s$^{-1}$.
In that case, we probably see a static structure, which will be 
dissolved after the dissolution time 
$t_{\mathrm{diss}} \sim r_{\mathrm{sh}}/\sigma_{\mathrm{ISM}}$, 
where $r_{\mathrm{sh}}$ is the size of the shell.

\subsection{Verification using the central pixel spectrum} 

As a third step in the identification scheme the spectrum through the center 
of the structure is automatically checked. The coordinates of the 
center ($l_{\mathrm{c}}$, $b_{\mathrm{c}}$, $v_{\mathrm{LSR}}$) are mean 
values of all pixels forming the HI shell candidate. A 
$\Delta T_{\mathrm{B}} = T_{\mathrm{hole}} - T_{\mathrm{bg}}$ spectrum through
the center of an \hi shell candidate is analyzed for depressions and peaks.
The background brightness temperature $T_{\mathrm{bg}}$ is defined in each 
velocity channel as an average emission from a strip around the structure with
$b=b_{\mathrm{c}}$. This definition proved to be more satisfactory than other
tested methods, e.g., the average emission from the surroundings of the 
structure.

\begin{figure}
 \centering
 \includegraphics[angle=270,width=8.0cm]{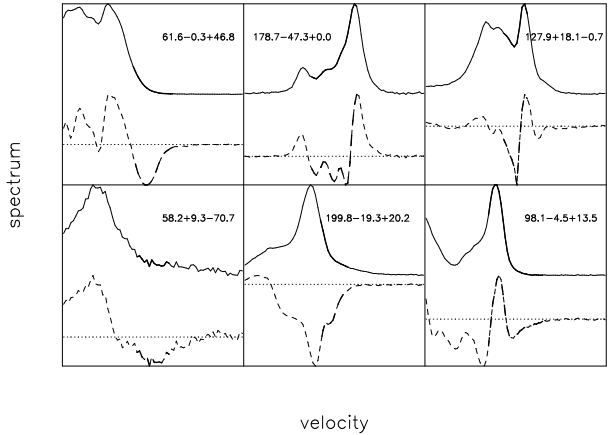}
 \caption{Spectra through the central pixel of \hi shell candidates. 
          The upper row shows spectra where the presence of an \hi hole 
          is confirmed. The lower row shows spectra where the signature of 
          an \hi hole is not clear or not present. Solid lines are observed 
          $T_{\mathrm{B}}$ spectra, dashed lines are $\Delta T_{\mathrm{B}}$ 
          spectra, a dotted line shows the background level. Velocity 
          channels in which the \hi shell candidate (steps 1 and 2) are 
          detected are plotted by the thick line. Numbers are coordinates 
          of structures ($lbv$).}
 \label{figspec}
\end{figure}

Structures that do not contain a clear depression in the
$\Delta T_{\mathrm{B}}$ spectrum or for which the depression is not 
located  in the velocity interval where the hole is visible in $lb$ maps 
are excluded from further study. Fig. \ref{figspec} shows examples of spectra.
Structures in the upper row are identified in velocity channels where the 
$\Delta T_{\mathrm{B}}$ spectrum has a minimum, therefore all are \hi holes. 
The upper left gives the spectrum showing a minimum in those velocity 
channels where the HI shell candidate is identified (thick dashed line). 
The upper center shows a  more complex spectrum with several smaller peaks 
in velocity channels corresponding to a temperature depression of the 
HI shell candidate. They probably come from small cloudlets engulfed by the 
HI hole. The upper right gives an example of a spectrum with a wall, which 
closes the HI hole at the position of the central pixel. On the other hand 
the three spectra in the lower row do not clearly demonstrate an 
\hi structure. The spectrum in the velocity channels where the HI shell 
candidate resides is either too flat (left) or the identified structure is 
displaced from the minimum  in $\Delta T_{\mathrm{B}}$ (center) or the 
structure is not connected to the minimum in  $\Delta T_{\mathrm{B}}$ 
(right).  

The interval $\Delta v$ between peaks surrounding the depression in 
$\Delta T_{\mathrm{B}}$ or between endpoints of upward slopes if peaks do 
not exist, equals twice the expansion velocity of the structure 
$v_{\mathrm{exp}} = {1 \over 2}\Delta v$. The $\Delta v$ interval is usually 
larger than the velocity interval where the hole is visible in $lb$ maps.  
This is because the algorithm searches for holes (i.e. minima in 
$\Delta T_{\mathrm{B}}$) and moreover for holes with dimensions larger than 
a given size. Therefore, velocity channels that contain low-contrast or 
small parts of the structure are not included in an \hi shell candidate 
created in the first two steps of the identification. Only a minority of 
structures has an interval $\Delta v$ from the spectrum analysis smaller 
than the visibility interval. This is usually because several substructures 
exist inside the hole, e.g. interacting shells, or preexisting \hi clouds. In 
such a case it is difficult to decide which peak or upward slope corresponds 
to the rim of the structure.
 
\subsection{Tests on artificial data}

Tests with artificial data show that completeness in identification of 
high-contrast shells is good regardless of gradients in the background. 
Low-contrast shells in the rapidly changing background can be misidentified 
as being smaller, or are completely missed. Small shells (compared to 
the thickness of the disc) are more easily identified than large ones. Random 
brightness temperature fluctuations create structures with a small velocity 
extent, which are easily removed when checking the reality of the feature
by analyzing a spectrum taken through the central pixel.  

When extrapolating these results from artificial data to types
of real shells we expect that the completeness of our identification for 
younger (i.e. not blown-out and not very distorted) shells is quite high, 
while older shells, which are blown-out and have fragmented walls, can be 
missed. We cannot detect shells that are formed by incomplete arcs: we 
detect only structures with a hole inside or completely encircled regions.

\section{Other automatic searches for \hi shells}

Another approach in the search for \hi shells uses models of expanding 
structures (usually hydrodynamic simulations) and then it searches in 
the observed data cubes for patterns similar to artificial data from  models. 
This method was applied by Thilker et al. (1998) and Mashchenko et al. (1999) 
for \hi shells in the galaxy NGC 2403, or by Mashchenko \& St-Louis (2002), 
who use a simple analytical model to search for \hi counterparts of a few 
\hii regions in CGPS. Calculated shapes of \hi shells are usually quite 
smooth and regular, due to a smooth density distribution in the 
model. More complicated structures with irregular walls and noisy velocity 
patterns, which are results of the turbulent nature of the ISM and of 
preexisting \hi structures, remain undiscovered by this method. Our approach 
does not use any precomputed shapes and  it can discover rather complicated 
forms. The only limitation is the angular and velocity resolution and the 
sensitivity limit.

The second approach primarily uses dynamical characteristics of expanding 
shells (Daigle et al. 2003). It searches for a signature of an expansion in 
the velocity spectra, i.e. the existence of two peaks separated approximately 
by twice an assumed expansion velocity. Regions with these features are then 
tested if their appearance corresponds to an assumed shape of shells. This 
method was applied to a small field of CGPS data, where it found two expanding
bubbles.

In our method the spectrum through the center of an \hi shell candidate is 
tested after the region is chosen as a depression in $lb$-maps, 
which is one difference between the methods. The other difference  is that 
we do not prefer any expansion velocity and we do not require the existence 
of peaks in a spectrum, which is a more general condition. We do not put any 
restrictions on the shape. The spectrum analysis serves as a way to 
distinguish between random configurations of pixels and physical structures. 

The method as a whole is quite robust. It is not sensitive to the adopted 
model of a shell, as is the first approach described above, and the spectrum 
through the structure need not contain peaks, as is the prerequisite of the 
method used by Daigle et al. (2003). Our approach resembles the
reversed method of de Heij et al. (2002) used for identification of high 
velocity clouds in LDS.

The third --- but from the chronological point of view the first --- method
is a traditional identification  by eye. With the growing amount of data and 
increasing resolution this method becomes more difficult and more 
time-consuming. Visual identification, compared to automatic methods, is not 
very effective for smaller and more regular or semi-regular shells. It is, 
however, so far unsurpassed when it comes to very irregular, patchy, 
non-expanding and open structures because of the ability 
of the human eye to combine disconnected features into a single shape.

\section{\hi shells in the Leiden-Dwingeloo survey}

\begin{figure*}
 \centering
 \includegraphics[angle=270,width=17.0cm]{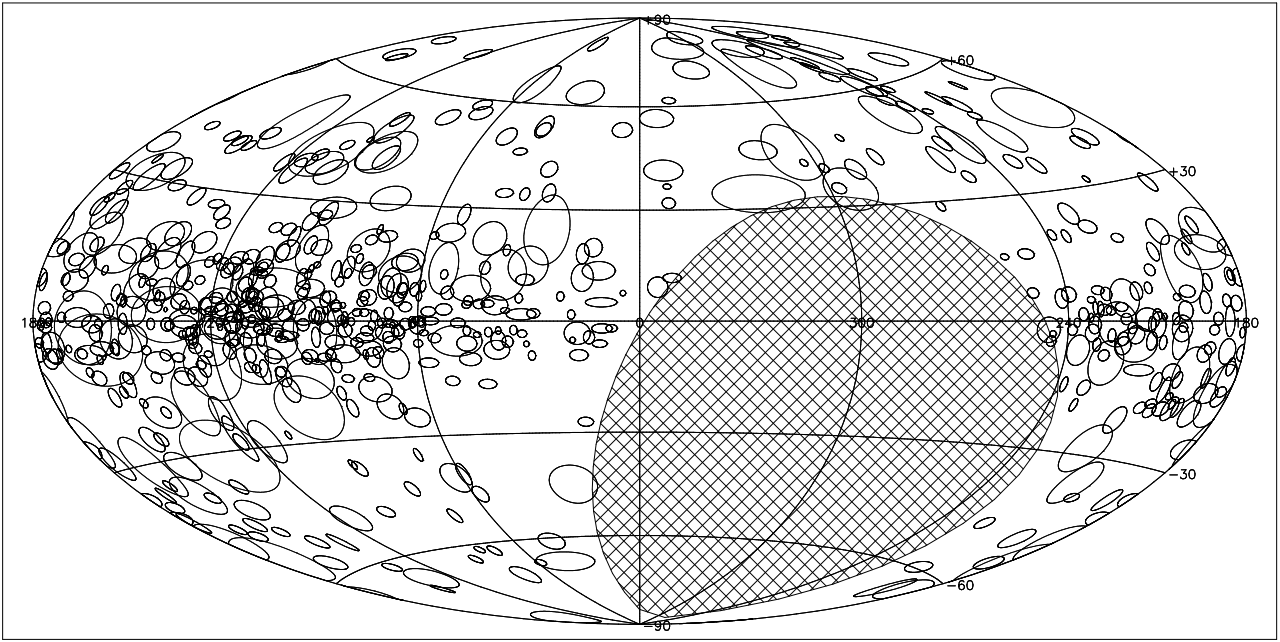}
 \caption{Positions of identified \hi shells in the LDS. The cross-hatched 
          region around $l=300^{\circ}$ is a region not accessible to the 
          Dwingeloo telescope.}
 \label{figpose}
\end{figure*}

In the whole LDS we have identified more than 600 structures that passed 
the three steps in our identification scheme. Their positions on the sphere 
are shown in Fig. \ref{figpose} and  their list is given in Table 1, 
which is available at the CDS. The table contains the following information: 
column 1 -- the shell number,
column 2 -- the shell name in the form GSlll$\pm $bb$\pm $vvv, 
column 3 -- the galactic longitude of the center $l_{\mathrm{c}}$ 
            in degrees, 
column 4 -- the galactic latitude of the center $b_{\mathrm{c}}$
            in degrees, 
column 5 -- the radial velocity of the center relative to the local 
            standard of rest $v_{\mathrm{LSR}}$
            in km s$^{-1}$, 
column 6 -- the shell dimension in the galactic longitude
            direction $\Delta l$ in degrees, 
column 7 -- the shell dimension in the galactic latitude
            direction $\Delta b$ in degrees, 
column 8 -- the shell extent in radial velocity $\Delta v$
            in km s$^{-1}$, 
column 9 -- a quality index, 
column 10 -- a corresponding structure identification in other lists of \hi 
             shells. Notes (1) and (2) in column 10 are explained in 
             the {\tt ReadMe} file of the Table.
The quality index is based on the visual appearance of the structure and it 
is not part of the automatic search. A quality 
index of 1 denotes clearly visible and pronounced shells, while an index 4  
refers to low-contrast and not well-pronounced shells. Indices 2 and 3 
are intermediate steps. 

There is substantial overlap of shells in the inner galaxy, which causes 
incompleteness of the identification. To avoid its influence on the results 
we decided to restrict further analysis to 276 \hi shells discovered in the 
2nd galactic quadrant. Moreover, the 2nd quadrant is fully covered in LDS. 

\begin{figure}
 \includegraphics[angle=270,width=8.0cm]{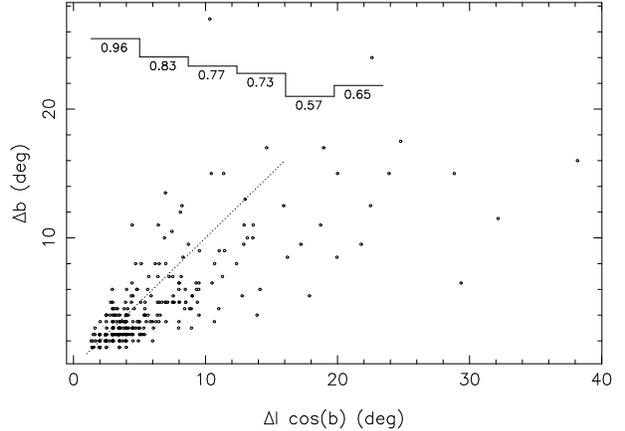}
 \caption{Angular dimensions $\Delta b$ against $\Delta l\  cos b$
          of shells identified in the 2nd Galactic quadrant. The dotted 
          line corresponds to $\Delta b = \Delta l\ cos b$. The upper solid 
          line gives the average values of $\Delta b / \Delta l\ cos b$ in 
          $\Delta l\ cos b$ bins.}
 \label{figangs}
\end{figure}

\begin{figure}
 \includegraphics[angle=270,width=8.0cm]{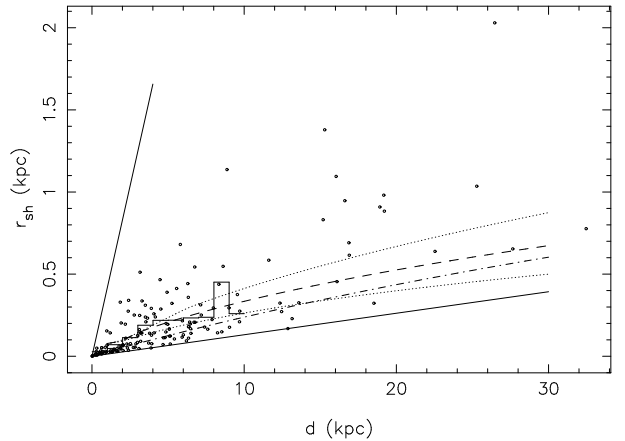}
 \caption{Radius of the shells $r_{\mathrm{sh}}$ versus their kinematic
          distances. The two solid lines give the minimum and maximum 
          $r_{\mathrm{sh}}$ corresponding to $1.5^\circ /2 $ and 
          $45^\circ /2 $ at 
          the distance $d$. Dashed, dotted and dash-dotted lines give the 
          average $r_{\mathrm{sh}}$ calculated for different models of size 
          distribution of shells given by eq.(2). The histogram gives 
          the average radius $r_{\mathrm{sh}}$ of the observed shells in 
          separate distance bins.} 
 \label{figdistsize}
\end{figure}

The majority of structures have $\Delta l\ cos b > \Delta b$ (see Fig. 
\ref{figangs}), although there is a population of $b$-elongated 
structures (upward and to the left of the dotted line in 
Fig. \ref{figangs}). Some of them are worms. All large shells 
($ > 15^{\circ}$) are, with one exception, more elongated in the $l$ 
direction than in the $b$ direction. This can partly be a result of the 
incomplete identification: because of the blowout effect the large shell 
is open to high $z$, and only a part of the structure is identified. This 
leads to an artifically lower $\Delta b$.

The ratio $\Delta b / \Delta l cos(b)$ is a good estimate of the shape
of shells parallel or perpendicular to the Galactic equator. This 
ratio is not a reliable parameter for very irregular shells or shells
elongated in directions other than the $l$ or $b$.

We calculated kinematic distances $d$ using the Wouterloot at al. (1990) 
rotation curve. Some shells have forbidden velocities or lie close to the 
anticenter direction. This reduces the number of shells in the 2nd quadrant 
with kinematic distances to 168. The angular dimensions
$\Delta l\ cos b,\ \Delta b$ are transformed to linear sizes. The shell radius
$r_{\mathrm{sh}}$ is given as 
\begin{equation}
   r_{\mathrm{sh}} = 
   {1 \over 4} d\  [tan (\Delta l\ cos b) + tan (\Delta b)].
   \label{eqrshcalc}
\end{equation} 

The dimensions of structures as a function of heliocentric distance are 
shown in Fig. \ref{figdistsize}. The solid lines indicate the limits 
imposed on $r_{\mathrm{sh}}$ by restricting the range of angular sizes of 
the shells to the interval (1.5, 45) degrees. Let us assume
that the size distribution of shells has a universal power-law form 
(Oey \& Clarke 1997):  
\begin{equation}
   dN(r_{\mathrm{sh}}) \propto {r_{\mathrm{sh}}}^{-\alpha} d r_{\mathrm{sh}},
   \label{eqsizedist}
\end{equation}
where $dN$ is the number of shells with radius  
$r_{\mathrm{sh}} \in (r_{\mathrm{sh}}, r_{\mathrm{sh}} + d r_{\mathrm{sh}})$.
For shells with the size distribution (\ref{eqsizedist}) we calculate the
expected average size for a given heliocentric distance $d$ as
\begin{equation}
   \bar r_{\mathrm{sh}} = {\int_{r_{\mathrm{min}}}^{r_{\mathrm{max}}}
   r^{-\alpha + 1} dr 
   \over {\int_{r_{\mathrm{min}}}^{r_{\mathrm{max}}}
   r^{-\alpha } dr }}.
   \label{eqrshave}
\end{equation} 
Integration ranges $r_{\mathrm{min}}$ and $r_{\mathrm{max}}$ are given
either by real physical limits (minimum and maximum possible radius of
the shell) or by the resolution limits 
($d \ tan 1.5^{\circ} /2, d\ tan 45^{\circ} /2$) 
of the searching algorithm, whichever is appropriate (larger or smaller). 
The major part of the theoretical curves in Fig. \ref{figdistsize} uses
$r_{\mathrm{min}} = d \times tan 1.5^{\circ}/2$ and 
$r_{\mathrm{max}} = 1.3$ kpc (if not said otherwise; for the choice of
1.3 kpc see Oey \& Clarke, 1997). The thick dashed line in 
Fig. \ref{figdistsize} refers to the size distribution with $\alpha = 2$; 
the thick dot-dashed line to $\alpha = 3$. Thin dotted lines have 
$\alpha = 2$ and $r_{\mathrm{max}}$ = 2.6 kpc (the upper line) or
$r_{\mathrm{max}}$ = 0.65 kpc (the lower line). According to a 
comparison with the histogram in Fig. \ref{figdistsize}, showing the average 
observed shell radii in separate distance bins, it seems that the power-law 
with $\alpha = 2$ and the maximum radius of the shell 
$r_{\mathrm{sh}} = 1.3$ kpc best fits the observations, but the spread  
in $r_{\mathrm{sh}}$ is large.   

\begin{figure}
 \includegraphics[angle=270,width=8.0cm]{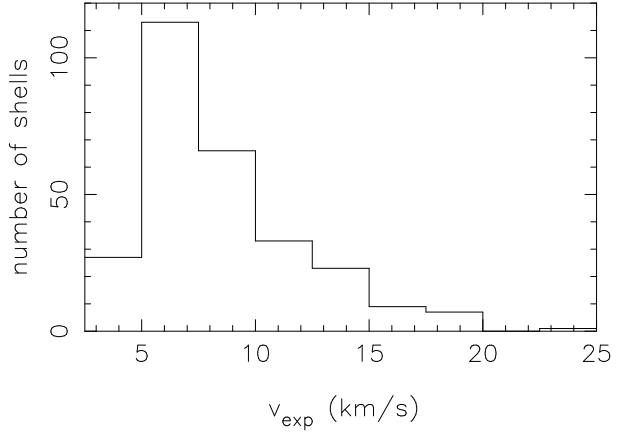}
 \caption{Expansion velocities of identified \hi shells in the 2nd 
          Galactic quadrant.}
 \label{figvexp}
\end{figure}

The distribution of expansion velocities  
$v_{\mathrm{exp}} = {1 \over 2} \Delta v$ is shown in Fig. \ref{figvexp}.
The largest observed expansion velocity is $25\ \mathrm{kms^{-1}}$; 
the median value lies between 5 and 10 $\mathrm{kms^{-1}}$. There are 
only a few structures with $v_{\mathrm{exp}} < 5\ \mathrm{kms^{-1}}$, which is 
apparently because such non-expanding structures  cannot survive for very long 
in a turbulent ISM (see section 3.2).

\subsection{Radius vs expansion velocity}

\begin{figure}
 \includegraphics[angle=270,width=8.5cm]{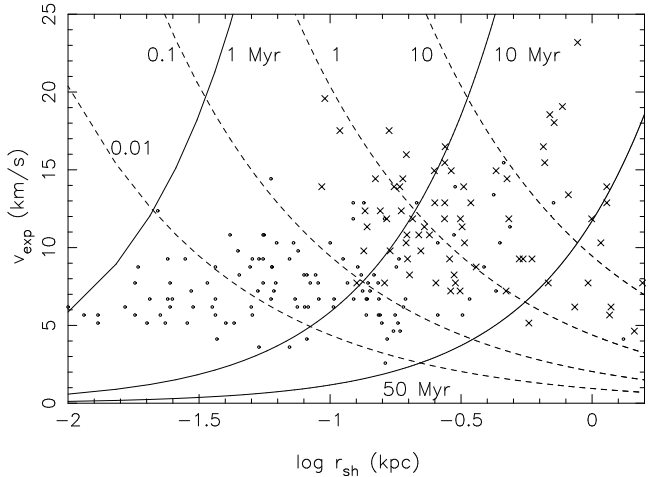}
 \caption{Radius of shell vs. expansion velocity for shells in the 2nd
          quadrant. Lines are analytical solutions after Weaver et al. (1977).
          Solid lines are lines of constant expansion time (1, 10, 50 Myr), 
          dashed lines are lines of constant $L/n$ (0.01, 0.1, 1, 10 
          $\mathrm{SN\ Myr^{-1}/cm^{-3}}$). Different symbols denote 
          different energies of shells (according to Chevalier, 1974), 
          circles are shells with $E < E_{\mathrm{SN}}$, crosses are shells 
          with $E \ge E_{\mathrm{SN}}$.}
 \label{figrv}
\end{figure}

A relation between the shell radius $r_{\mathrm{sh}}$  and the expansion 
velocity $v_{\mathrm{exp}}$ is shown in Fig. \ref{figrv}. We also plot lines 
of constant expansion time and luminosity over the ISM density ($L/n$)
according to an analytical solution of Weaver et al. (1977). As expected, 
there are no shells younger than 1 Myr and there are 
only a few shells with ages greater than about 50 Myr. Young shells are small 
and not fully developed, while old shells are either completely destroyed by 
external forces (like spiral arms) or too fragmented for our method to 
identify. There is a lack of small shells with high expansion velocities, but 
this is not surprising. Shells smaller than about 50 pc are either 
wind-blown bubbles or remnants after one supernova explosion. Such objects 
tend to have expansion velocities around or lower than 
$10\ \mathrm{kms^{-1}}$. Dimensions of real ``supershells'', created by 
several OB stars in associations, start at radii around 100 pc. Some 
supershells have expansion velocities larger than  $10\ \mathrm{kms^{-1}}$, 
though we do not see any clear dependence of radius on velocity. The argument 
that small shells are wind-blown bubbles and large shells are created by 
several SNs is further supported by their energies (Chevalier, 1974). All 
small shells have energies lower than the typical energy 
of one supernova explosion $E_{\mathrm{SN}} = 10^{51} \mathrm{erg}$.

It is interesting to compare the properties of shells in the Milky Way with 
those in external galaxies. Walter \& Brinks (1999; further WB99) made a 
comparison of four galaxies, M31, M33, IC 2574 and Holmberg II. They plot 
a graph of radius vs. expansion velocity (Fig. 20 in their paper) for these 
four external galaxies. They find that the two spiral galaxies (M31 and M33)
have on average smaller shells and a larger range of expansion velocities 
compared to the dwarfs IC 2574 and Ho II. Neither dwarf contains
relatively small shells (radius around 100 --- 300 pc) with high expansion 
velocities (around $20\ \mathrm{kms^{-1}}$), while there is a significant 
population of such shells in M31 and M33. The Milky Way behaves like a spiral 
galaxy in this aspect; there are many expanding shells with a 
100-300 pc size. Contrary to M31 and M33, in the Milky Way there are more 
shells with radii greater than 800 pc, while there is only one in M31 and 
a few in M33. This can be partly due to a different inclination angle of 
galaxies, because different viewing angles may emphasize different 
parts of shells. This effect is only important for non-spherical, 
therefore predominantly large shells.

Differences in the $r_{\mathrm{sh}}$ vs $v_{\mathrm{exp}}$ graph can be 
explained by different conditions between spiral and dwarf galaxies and by 
different star formation histories. It is possible to calculate the age of 
a shell from its size and expansion velocity, e.g. using the solution
of Weaver et al. (1977), see Fig. \ref{figrv}. According to WB99, spiral 
galaxies M31 and M33 have many young shells (younger than 10 Myr), but nearly 
no old ones (older than 30 Myr), while dwarf galaxies IC 2574 and especially 
Ho II have few young shells, but many old ones. The mean ages of shells in 
these galaxies are 6 Myr (M31), 7.2 Myr (M33), 14 Myr (IC 2574) and 37 Myr 
(Ho II); all quoted values were multiplied by a factor of 3/5 to change them 
from the simple $r$/$v$ used in WB99 to the solution of Weaver et al. (1977). 
The differences are explained by different star formation histories and a 
higher rate of shell destruction by density waves in spiral arms. 

A somewhat different behavior was found for the Magellanic Clouds. 
Kim et al. (1999) shows an $r_{\mathrm{sh}}$ vs $v_{\mathrm{exp}}$ graph for 
the LMC. There is a population of young shells, but there are no shells larger 
than 600 pc. This is not caused by the missing short spacing data, 
as it has been shown by Kim et al. (2003) and Staveley-Smith et al. 
(2003): there were no new identifications of supershells when the Parkes and 
ATCA observations were combined. Shells larger than 600 pc are 
intrinsically missing in the LMC. The mean age of shells in the LMC is 
4.9 Myr.

This is similar to the mean age of shells in the SMC (Staveley-Smith et al. 
1997), 5.4 Myr. The list of shells in the SMC also suffers from the 
lack of large shells, but this effect is not so severe: in the 
combined Parkes and ATCA data (Stanimirovi\`c et al. 1999) only three new 
supershells with $r_{\mathrm{sh}} > 600\ \mathrm{pc}$ were found. The behavior
of the Magellanic Clouds is clearly different from that of the dwarfs IC 2574 
and Ho II. The presence of young shells in the SMC is explained by significant
star formation induced by interaction with the Milky Way, but the lack of 
old shells -- all the SMC shells are younger than 30 Myr -- is somewhat 
puzzling since the last SF burst in the SMC happened 200 Myr ago 
(Stanimirovi\`c et al. 2004).

The mean age of shells in our list is 8.4 Myr. This value is influenced 
by different selection effects, i.e. the detection of old shells is 
much less complete than of young ones since old ones are less continuous 
and more fragmented. All shells with ages greater than 30 Myr are found at 
galactocentric distances larger than 12 kpc --- one shell with an age of 
30 Myr is located at $R= 12\ \mathrm{kpc}$, the rest of the old shells are 
at distances greater than 16 kpc. The age distribution of shells is similar 
to that for spiral galaxies M31 and M33.

Our search algorithm is not very sensitive to fragmented structures, which
may explain the lack of old shells. However, as we do not see any old shells 
at distances smaller than 12 kpc, while there are some at larger distances, we 
can at least say that the lifetime of shells is longer at greater
galactocentric distances.

\subsection{\hi density}

\begin{figure}
 \includegraphics[angle=270,width=8.5cm]{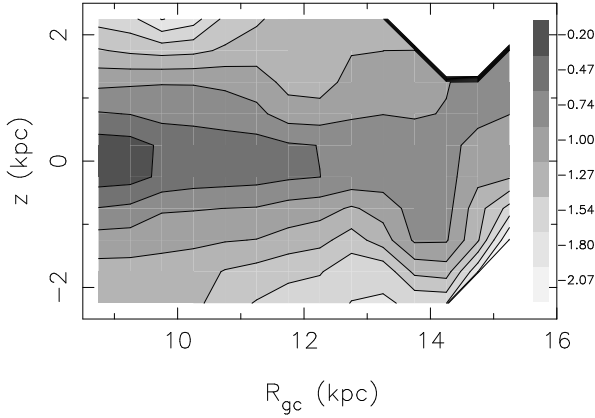}
 \caption{An average \hi density derived for shells at different positions 
          in the Galaxy. The grey scale is logarithmic - the darkest
          color corresponds to 0.63 cm$^{-3}$, lightest color to 0.01
          cm$^{-3}$, the  white color denotes positions with no information 
          about the density, i.e. without an \hi shell.
         }
 \label{figdens}
\end{figure}

\begin{figure}
 \includegraphics[angle=270,width=8.5cm]{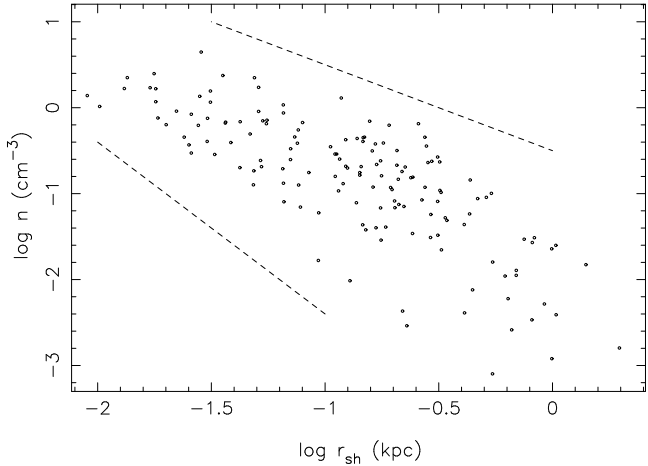}
 \caption{The \hi density versus size of the shells $r_{\mathrm{sh}}$. Two 
          dashed auxiliary lines are power-laws with indices of -1 and -2.
         }
 \label{rshn}
\end{figure}

The original \hi density at the place of a shell is calculated  as the
density corresponding to the mass missing in the \hi hole. If that
mass were 
evenly added to the hole, the structure would disappear and the observed \hi 
column density would be equal to the average value around the shell.  
The maximum value found for an individual \hi shell is 
$24\ \mathrm{cm^{-3}}$ for a very small nearby shell and 
$2.2\ \mathrm{cm^{-3}}$ for shells with a radius greater than 
$50\ \mathrm{pc}$. The typical density is around $0.5\ \mathrm{cm^{-3}}$. 
We calculated an \hi density $n$ in positions of all shells 
and made an average for each volume element of the galactic disc
(Fig. \ref{figdens}). As expected, the density decreases with increasing 
$|z|$ and $R$ coordinates. These reflect the density gradients in the HI 
galaxy disk in combination with the shell size versus density effect: the 
observations show that shells expand more in a lower density medium. This 
is demonstrated in Fig. \ref{rshn}, where we plot sizes of shells 
$r_{\mathrm{sh}}$ versus densities $n$.

\subsection{Energy}

\begin{figure}
 \includegraphics[angle=270,width=8.5cm]{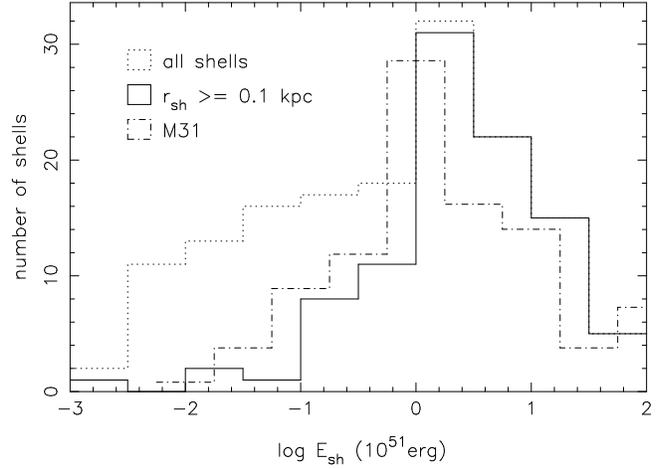}
 \caption{The distribution of energy deposited in \hi shells. The 
         distribution of all shells (dotted line) and the distribution of 
         shells with size $r_{\mathrm{sh}} > 0.1$ kpc (solid line). The 
         distribution for the M31 galaxy (dash-dotted line) is taken from 
         Walter \& Brinks (1999); it is scaled to the Milky Way values.
         }
 \label{figener}
\end{figure}

The most widely used method to estimate the energy of \hi shells is 
Chevalier's formula (Chevalier 1974),
\begin{equation}
    {E_{\mathrm{tot}} \over \mathrm{erg}} = 5.3 \times 10^{43}
    \left ({n \over \mathrm{cm}^{-3}} \right )^{1.12}
    \left ({r_{\mathrm{sh}} \over \mathrm{pc}} \right )^{3.12}
    \left ({v_{\mathrm{exp}} \over \mathrm{km s}^{-1}} \right )^{1.4}
    \label{eqcheval}
\end{equation}
where $n$ is the density of the ambient medium, $r_{\mathrm{sh}}$ the
radius of the shell and $v_{\mathrm{exp}}$ its expansion velocity.
We take $E_{\mathrm{SN}} = 10^{51}\mathrm{erg}$. Energy estimations using 
Eq. (\ref{eqcheval}) of smaller and less evolved shells are quite reliable, 
but for non-spherical shells it gives an underestimation of the energy  
that created the structure. 

The distribution of the energies in \hi shells according to the Eq.
(\ref{eqcheval}) is shown in Fig. \ref{figener}. There are many low-energy 
structures ($E < 0.1\ E_{\mathrm{SN}}$) that were probably not created 
by SN explosions. The majority of them are small with radii 
$r_{\mathrm{sh}} < 100\ \mathrm{pc}$. Many of them are wind--blown 
shells. Even though Chevalier's Eq. (4) has the largest 
dependence on radius, many large shells do not belong to the most 
energetic structures since the density $n$ is systematically smaller for large
compared to small shells; this influences the energy derived with 
Eq. (\ref{eqcheval}) in the opposite direction to the size $r_{\mathrm{sh}}$. 

There are only 12 \hi shells with an energy 
$E > 20\ E_{\mathrm{SN}}$ and one with an energy exceeding 100 
$E_{\mathrm{SN}}$ (its energy is 120 $E_{\mathrm{SN}}$). 
This can be understood because we study the outer Milky Way and we do 
not observe many energetic sources there, at least not many OB associations.
Bureau \& Carignan (2002) suggest that ram pressure due to a galaxy 
moving through the inter--galactic medium of a group or cluster can remove 
material from the edges of the holes and prevent it from refilling the 
holes' interiors. Thus, ram pressure enlarges preexisting holes and 
lowers the amount of energy necessary for their creation. Ram pressure and 
high velocity cloud impacts apply particularly at the galactic peripheries 
where the gas density and star formation rates are low.

We over-plot a scaled energy distribution of shells in M31 (a dash-dotted
line in Fig. \ref{figener}) taken from WB99. They found that the distribution
is very similar in the four galaxies (M31, M33, IC 2574, Ho II).  The Milky 
Way distribution is similar as well: it has a maximum at the same value 
(around $1\ E_{\mathrm{SN}}$) and the same decline for high energies. However,
we detect more low-energy shells than were found in M31, which may be partly 
a selection effect due to the larger distance to M31, limiting the linear 
resolution. Another explanation is that the Milky Way creates more of them, 
or because our search algorithm is more sensitive than a visual inspection,
upon which the M31 distribution (and all others in WB99) is based. The mean 
energy of shells in M31 is $15\ E_{\mathrm{SN}}$, similar to M33 
($16\ E_{\mathrm{SN}}$), IC 2574 ($11\ E_{\mathrm{SN}}$) and Ho II 
($14\ E_{\mathrm{SN}}$). The mean energy of shells in the outer Milky Way is 
$3\ E_{\mathrm{SN}}$ ($4.7\ E_{\mathrm{SN}}$ 
for shells with $r_{\mathrm{sh}}>100\ \mathrm{pc}$), this lower value is 
related to the extended low-energy wing in the energy distribution function.

\section{Distribution of shells in the Milky Way from Distance Effect Removed 
(DER) samples} 

\begin{figure*}
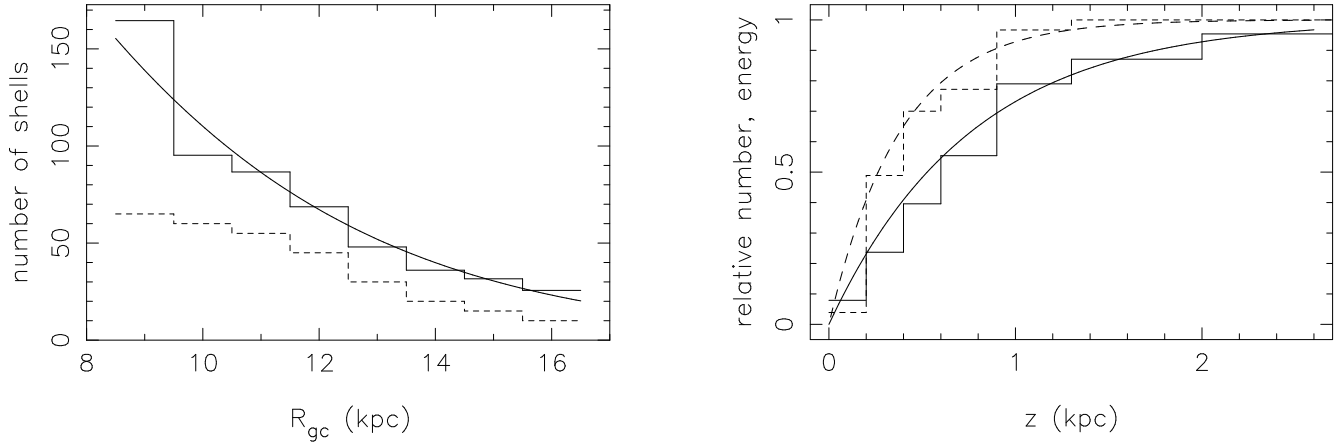

 \includegraphics[angle=270,width=8cm]{figmwshell12a.ps}
 \hskip 1.5cm 
 \includegraphics[angle=270,width=8cm]{figmwshell12b.ps}
 \caption{The radial and $z$-distribution of \hi shells. The left panel
          shows recomputed numbers of shells derived from a DER sample 
          by multiplication with a geometric factor (see the text)
          at different galactocentric radii and an exponential fit with 
          $\sigma_{\mathrm{gsh}} = 3$ kpc (solid line). 
          The dashed line gives 
          the numbers from the DER sample (see the text). The right 
          panel is the $z$-distribution of \hi shells. Solid lines show the 
          normalized cumulative number of shells at coordinates $z$ lower 
          than a given value, dashed lines show the normalized energy in 
          these shells. Binned data are observations, lines are exponential 
          profiles with a 
          scale length of 760 pc (solid) and 380 pc (dashed).
          Corresponding HWHMs are 530 and 270 pc.}
 \label{figradz}
\end{figure*}

\subsection{Distance Effect Removed (DER) samples}

Due to our position inside the Milky Way our list of shells suffers
from various distance--dependent selection
effects. Intrinsically small structures are seen
only in the vicinity of the Sun. When they are further out, they are too 
small to be either observed by the telescope or detected by our algorithm. 
On the other hand, large shells, when close to the Sun, are angularly extended
and such objects are difficult to identify because they are too big for the 
field in which identifications take place, because of intervening material, 
and sometimes also because we usually see only walls of these structures and 
they are incomplete or fragmented. To avoid these problems we create 
'distance effect removed (DER) samples', where we include shells with 
dimensions and distances such that their angular dimensions fall into an 
interval
$(\phi_{\mathrm{min}},\phi_{\mathrm{max}})=(1.5^{\circ};\ 15^{\circ})$. 

The lower limit $\phi_{\mathrm{min}}$ is the minimum dimension for 
identification, the upper limit $\phi_{\mathrm{max}}$ is smaller than the 
upper limit for identification (45$^\circ $). This choice  was made 
to ensure that we do not miss any structure in the selected interval. The 
range of heliocentric distances (or shell dimensions) is a free parameter.
When processing a DER sample, we work with shells whose linear dimension
$\Delta _{\mathrm{sh}} = 2 r_{\mathrm{sh}}$ is $\Delta _{\mathrm{sh}}
\in (\Delta _{\mathrm{min}},\Delta _{\mathrm{max}})$ and  heliocentric 
distances $d$ are $d \in (d_{\mathrm{min}},d_{\mathrm{max}})$, where 
$\Delta _{\mathrm{min}} = d_{\mathrm{max}}tan(\phi_{\mathrm{min}})$ and 
$\Delta _{\mathrm{max}} = d_{\mathrm{min}}tan(\phi_{\mathrm{max}})$. With this
we ensure that the smallest shell detected at the largest distance 
$d_{\mathrm{max}}$ has an angular size equal to $\phi_{\mathrm{min}}$
and at the same time that the largest shell at the smallest distance 
$d_{\mathrm{min}}$  has an angular size equal to $\phi_{\mathrm{max}}$.  
We analyze different DER samples selected with different choices of
minimal and maximal heliocentric distances $d_{\mathrm{min}},d_{\mathrm{max}}$.
All shells in studied DER samples are larger than $100\ \mathrm{pc}$, 
which avoids contamination by wind-blown bubbles and low-energy structures,
and smaller than 700 pc to avoid an incompletness in identifications of
large evolved shells.

\subsection{Radial distribution}

It is generally thought that \hi shells are mainly produced by SN explosions 
and other energetic activities in OB associations. Therefore their radial 
distribution should reflect the radial distribution of young stars and OB 
associations. Attempts to find a correlation between these distributions 
were made by Ehlerov\'a \& Palou\v{s} (1996) for the Milky Way Galaxy and
Palou\v{s} \& Ehlerov\'{a} (1997) for M31 and Holmberg II. The distribution 
of \hi shells in the Milky Way was constructed from the list of Heiles (1979) 
and an exponential decrease of $\Sigma _{\mathrm{gsh}} \propto 
exp^{-({R \over \sigma_{\mathrm{gsh}}})}$  was found, with 
$\Sigma_{\mathrm{gsh}}$ being the surface density of \hi shells, $R$ the 
galactocentric distance and $\sigma_{\mathrm{gsh}} \simeq 4.5\ \mathrm{kpc}$ 
the radial scale. However, the effects of heliocentric distance were not taken 
into account properly. In this paper we will perform a similar analysis on our 
new and better-processed list of \hi shells.

Another attempt to determine the galactic distribution of shells was made by 
McClure-Griffiths et al. (2002). Their list consists of shells from Heiles 
(1979) and their own identifications. They are mostly interested in the 
relation between shells and spiral arms. 

We fit an exponential profile to the radial distribution of shells identified 
and described in this paper:
\begin{equation}
    \Sigma _{\mathrm{gsh}}(R) = 
    \Sigma _{\mathrm{0gsh}} e^{-{R \over \sigma_{\mathrm{gsh}}}},
    \label{eqradial1}
\end{equation}
where $R$ is the galactocentric distance and $\sigma_{\mathrm{gsh}}$ is 
the radial scale length. The number of shells $dN$ at a given distance $R$ is 
\begin{equation}
   dN(R) = 2 \pi R \Sigma _{\mathrm{gsh}}(R)dR
    \label{eqradial2}
\end{equation}

The observed numbers of shells in intervals of galactocentric distances are 
multiplied by a geometric factor proportional to the ratio of the surface 
of the whole galactocentric ring and the surface of that part of the ring 
that lies in the studied heliocentric range. These recalculated numbers, 
which correspond to $dN(R)$ in Eq. (\ref{eqradial2}), are then fitted to the 
profile (\ref{eqradial1}). The resulting radial scale length 
$\sigma_{\mathrm{gsh}}$ for the DER sample with 
$\Delta _{\mathrm{min}}$ = 200 pc and $\Delta _{\mathrm{max}}$ = 800 pc is
\begin{equation}
   \sigma_{\mathrm{gsh}} = {3^{+2}}_{\hspace*{-0.35cm}-1}\ \mathrm{kpc}. 
   \label{eqsigmares}
\end{equation}
We exclude shells with $r_{sh} >$ 400 pc because
DER samples including these large shells would have $d_{\mathrm{min}}
> 3$ kpc. It would substantially reduce the number of shells in the
interval of galactocentric distances $R_{\mathrm{GC}} \simeq (9, 12)$ kpc and
increase the error of $\sigma_{\mathrm{gsh}}$. The fit, together with the 
observed and recalculated numbers of shells is shown in Fig. \ref{figradz} 
(left panel).

The radial scale length of about 3 $\mathrm{kpc}$ is comparable to that  
of the stellar disc. Luminosity models give values between 2.5-3.0 kpc 
(Kent, Dame \& Fazio 1991, $\sigma = (3 \pm 0.5)\ \mathrm{kpc}$; 
Freudenreich 1998, $\sigma = 2.5\ \mathrm{kpc}$). OB associations have a
radial scale length of 1.8 kpc (Bronfman et al. 2000). If the scale 
length of stars and \hi shells differed substantially, it would mean that 
an important part of the shells is not connected to the stellar activity. 
This,however, is not the case. The derived radial scale length of HI shells 
roughly corresponds to the stellar scale length and is larger than the 
scale length of OB stars. The scale length of OB associations is based mainly 
on measurements in the inner Galaxy 
(for $R_{\mathrm{GC}} \in (0.5;2.0)R_{\odot}$ 
with the majority of sources inside the solar circle), while the scale length 
of HI shells is derived only for the outer Galaxy. Therefore it is possible 
that the discrepancy is a result of a systematic difference between the inner 
and outer Galaxy. Another possible explanation is that a certain fraction of 
HI shells is created by a different mechanism than energy input from OB
associations.

\subsection{The distribution in the $z$-direction}

\hi shells are concentrated towards the Galactic plane. One half of all 
shells lie in a 1-kpc thin layer (the HWHM is around 500 pc). The thickness 
of this layer increases slightly with galactocentric distance: the HWHM 
for shells with $R \in (9,11)$ kpc is 530 pc (see Fig. \ref{figradz}),
for those with $R \in (12,14)$ kpc, it is 590 pc. A more precise determination
of this trend is not possible because of low number statistics
at larger galactocentric distances.

Identified \hi shells reside in a much thicker disc compared to the \hi disc 
or even to the stellar disc. It may be partly due to our identification 
procedure, which shifts the shell centers artificially to lower density ends, 
i.e. to higher $z$-coordinates. 

The average energy of shells (according to Eq. \ref{eqcheval}) decreases with 
increasing coordinate $z$ (see Fig. \ref{figradz}). It is uncertain if this 
is a real physical effect (fewer energetic sources at high $z$ 
coordinates) or simply a consequence of the fact that the volume
density decreases to higher values of $z$ and as a result the estimated 
energy is lower.   
The HWHM of the energy release is about half of that of the number density, 
i.e. around 250 pc, also increasing with increasing $R$, 200 pc for 
$R \in (9,11)\ \mathrm{kpc}$ and 230 pc for $R \in (12,14)\ \mathrm{kpc}$.

The exponential or Gaussian fit of the $z$-distribution of \hi shells is not 
as good as the radial one, but the exponential fit is better than 
Gaussian. The HWHM derived from these fits is larger than observed. This 
means that there are more shells at high coordinates than would correspond 
to an exponential (Gaussian) profile, i.e. the distribution of shells has 
extended wings. Fig. \ref{figradz} (right panel) shows the $z$-distribution 
of shells in the DER sample with 
$\Delta _{\mathrm{min}}$ = 200 pc and $\Delta _{\mathrm{max}}$ = 800 pc.

\subsection{Size distribution of shells}

Oey \& Clarke (1997; hereafter OC97) calculate the size distribution 
of shells from theoretical considerations of shell evolution. They found 
that the size distribution can be quite reliably approximated by a power law 
(\ref{eqsizedist}). They compare the size distributions of shells in the
galaxies M31, M33, Holmberg II and SMC and find that the power-law 
representation generally fits the data well with $\alpha \in (2.1, 2.7)$.

\begin{figure}
 \includegraphics[angle=270,width=8.5cm]{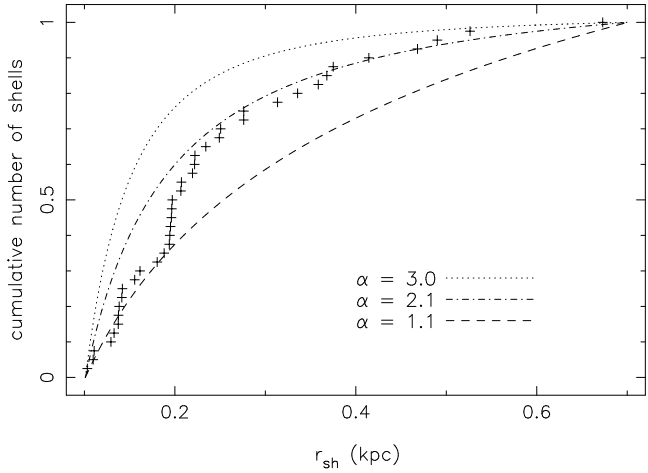}
 \caption{The size distribution of \hi shells. The number of shells with 
          radius larger than 100 pc and smaller than a given value of
          $r_{sh}$ vs. the radius $r_{sh}$ is plotted, plus 
          a theoretical value for three different power-laws.}
 \label{figsiz}
\end{figure}

Fig. \ref{figsiz} gives the size distribution together with three different 
power laws with $\alpha = 1.1, 2.1$ and 3.0. A fit of the DER sample
with $\Delta _{\mathrm{min}}$ = 200 pc and $\Delta _{\mathrm{max}}$ 
= 1 400 pc gives the average value of the index $\alpha$ around 2.1: 
\begin{equation}
   \alpha = (2.1 \pm 0.4). 
   \label{eqalphares}
\end{equation}
The spectrum seems to be flatter for smaller shells ($\alpha < 2.1$) and 
steeper for larger shells ($\alpha > 2.1$). There may be several 
regimes in the size distribution, one with $\alpha = 1.1$, one with 
$\alpha = 2.1$ and another with $\alpha = 3.0$. The difference in the 
indices may be real or it may be an artifact of the identification method. 
Evolutionary effects should lead to an increase of $\alpha$ for large shells: 
some of the large old shells are destroyed, fragmented or otherwise distorted 
and therefore are more difficult to find. Merging and grouping of small shells 
produces a change in $\alpha$ between small and intermediate shells: instead 
of several small shells we see one larger shell. Our identification method (as 
discussed earlier) can underestimate the dimensions of extended shells in 
the $z$ ($b$) direction. Indeed, Fig. \ref{figangs} shows that shells become 
more elongated in the $l$ directions with increasing $\Delta l$. This effect 
may also lead to higher $\alpha$ for large shells. An alternative explanation 
is that the completeness of our list is comparatively good for large shells, 
while it is worse for smaller shells, but this is unlikely as small 
shells are more regular and easier to find than large distorted structures. 

The power-law index of the luminosity distribution of energy sources 
$\Phi({\cal L})$ (e.g. OB associations) is given by OC97 as
\begin{equation}
   \Phi({\cal L}) \propto {\cal L}^{-\beta}, 
   \label{eqoey2}
\end{equation}
where ${\cal L}$ is the luminosity of the source. OC97 give the relation 
between the power laws (\ref{eqsizedist}) and (\ref{eqoey2}) as 
\begin{equation}
   \alpha = 2 \beta - 1,
\end{equation}
which would imply from our size distribution fit 
\begin{equation}
   \beta = (1.6 \pm 0.3).
   \label{eqbetares}
\end{equation}
However, the direct connection between sizes of shells and powering 
sources disregards the size-density relation shown in Fig. 10. 
It can be one of the systematic effects, which distinguishes between
different types of galaxies. The indices $\alpha$ and $\beta$ are on 
the shallower end of the corresponding values given by OC97 for M31, M33, 
Holmberg II and SMC; $\alpha \in (2.1, 2.7)$, $\beta \in (1.6, 1.9)$. The 
index $\beta$ is flatter than 2, which is the slope of the \hii 
region luminosity function (Kennicutt et al. 1989) or OB association 
luminosity function (McKee \& Williams 1997). This lower value may be 
connected to the blow-out effect in the thin \hi disc of the Milky Way. The 
sizes of shells powered by high ${\cal L}$ OB associations are smaller 
because a fraction of the energy leaks to high-$z$ distances. Blow-outs 
reduce the difference between the sizes of  \hi shells created by  high and 
low ${\cal L}$ OB associations. The shell size distribution then leads to 
an artificially flatter luminosity function 
than that of OB associations.

\section{Filling factor of \hi shells}

\begin{figure}
 \includegraphics[angle=270,width=8.5cm]{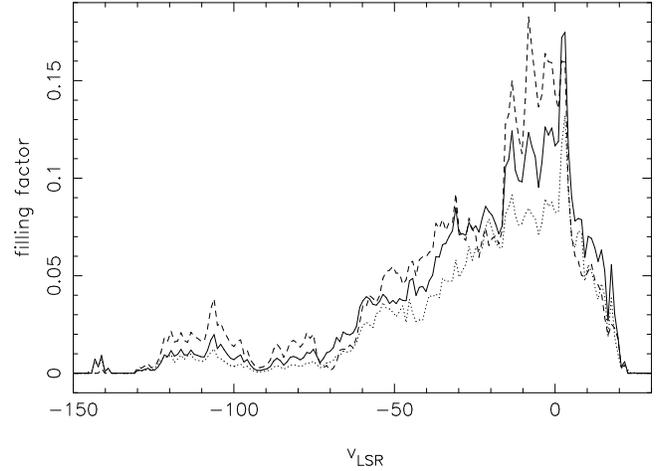}
 \caption{The 2--D filling factor of \hi holes in the  $T_B(l, b, v)$ 
          datacube for 
          $(l,b) \in (90^{\circ},180^{\circ};-10^{\circ},+10^{\circ})$
          (solid line), 
          $(l,b) \in (90^{\circ},180^{\circ}; -5^{\circ},+5^{\circ})$ 
          (dashed line), and 
          $(l,b) \in (90^{\circ},180^{\circ};-20^{\circ},+20^{\circ})$ 
          (dotted line) as functions of the radial velocity.}
 \label{figffv}
\end{figure}

The filling factor $f$ of HI shells is the total area $(f_{\mathrm{2D}})$ or 
volume $(f_{\mathrm{3D}})$ occupied by HI shells compared to the total area 
or volume of the galaxy. The derived filling factors are lower limits 
to intrinsic values, since some shells are missing (due to distance effects, 
incomplete identification, etc).

The 2--D filling factor as a function of radial velocity is shown in Fig. 
\ref{figffv}, where we show the results for strips along $b=0^{\circ}$.
The profiles for the wider or narrower intervals of $b$ are similar.
Neglecting distance effects, i.e. neglecting the fact that at different 
distances shells with different dimensions are not seen, the total filling 
factors in the 2nd galactic quadrant of the Milky Way are calculated. We get 
$f_{\mathrm{3D}} \simeq 0.05$ and $f_{\mathrm{2D}} \simeq 0.4$. 
We argue that neglecting distance effects we can approach to real, unbiased 
values. This is because filling factors are dominated by large shells, 
therefore the non-detection of small shells does not strongly influence 
the results.

\section{Summary}

We created an automatic method to search for \hi shells in data cubes. The 
method is based on finding local minima in individual velocity channels. 
These 2D regions are then combined along the third dimension, the radial 
velocity axis, to create 3D objects. The spectrum  through the center of the 
structure is analyzed. The method has the advantage of being independent of 
models of \hi shells and therefore being widely applicable. Small shells 
(compared to the thickness of the disc) are more easily identified than large 
ones. Random brightness temperature fluctuations create structures with a 
small velocity extent, which are easily discarded when looking at spectra 
taken through the centers of the shells. We cannot detect shells that are 
formed by incomplete arcs: we detect only structures with a hole inside or 
completely encircled regions. We detect more shells than published by Heiles 
(1979) and other authors. Some of the shells identified by us for the first 
time are as well defined as those previously catalogued.

In the whole Leiden-Dwingeloo survey we identify more than 600 structures.
To avoid problems connected with a substantial overlap of shells
in the inner Galaxy, we analyze only structures in the 2nd Galactic
Quadrant. There are nearly 300 shells found there. From their list we create 
samples that compensate for the distance effect, i.e. the fact that we see 
small shells only in the vicinity of the Sun, and that angularly large shells 
are difficult to identify in the data.

Firstly, we compare shells in the Milky Way with shells in external galaxies. 
We find that galactic shells have similar properties to shells in M31 and M33.
The majority of shells are $\sim$ 10 Myr old and there are few shells older 
than $\sim 30\ \mathrm{Myr}$. The energy distribution is similar to that of 
M31, but we detect relatively more low-energy shells.

Secondly, we study the Galactic distribution of \hi shells. 
The radial decrease in the surface density of shells in the outer
parts of the galactic disk is exponential, with a scale length 
of about 3 kpc. This is comparable to the radial scale length of the 
stellar disc, but larger than the radial scale length of OB associations. 
The similarity between scale lengths of stars and HI shells supports the 
idea that most shells are connected to stars. The reason for 
the discrepancy between scale lengths of \hi shells and OB associations is 
not clear, it may either reflect differences between the inner and
outer Galaxy, or it may result from the fact that a fraction of
shells in the outer Galaxy is created by a different mechanism than
by an energy input from young and massive stars.

\hi shells are concentrated towards the Galactic plane. Half of all shells 
lie in a 1-kpc thin layer (the HWHM is around 500 pc). The thickness of this 
layer increases slightly with galactocentric distance. The energy in shells 
is more concentrated toward the plane; half of all the energy is contained 
within the disc with a total thickness of 500 pc.

We then study the size distribution of \hi shells. We find that a
power--law function (\ref{eqsizedist}) with a slope $\alpha = 2.1 \pm 0.4$ 
describes the observed distribution. $\alpha$ seems to be larger for larger 
shells and smaller (flatter) for smaller shells. We attribute the increase 
of $\alpha$ with dimension to evolutionary effects and the inability of the 
identification method to find distorted and fragmented structures. The size 
distribution with $\alpha =2.1$ is flatter, but comparable to the 
distributions found in the galaxies M31, M33, Holmberg II and SMC. The 
corresponding slope of the luminosity function of sources powering \hi shells
is $ \beta = (1.6 \pm 0.3)$. This is also flatter than the observed 
distribution of \hii regions or OB associations in the Milky Way. 
However, the size of shells also depends on the density of the 
ambient medium. Consequently, part of the effect may be due to a 
size-density relation observed in our shell sample.

The filling factors are derived for shells indentified in the 2nd
galactic quadrant. Due to incompleteness of the sample, the derived
values $f_{\mathrm{3D}} =  0.05$ and $f_{\mathrm{2D}} = 0.4$ are lower limits 
to the intrinsic values. 
 
With $r_{\mathrm{sh}}$, $v_{\mathrm{exp}}$, the surface density of the shell
$\Sigma_{\mathrm{sh}}$ and with the speed of sound in the shell, it would be 
possible to estimate the gravitational stability of the shell and find out 
if it can trigger star formation. Some of the shells detected in the LMC, like
the Sextant region near LMC IV (Efremov et al. 1999), probably triggered 
star formation. A discussion of shell triggering and a comparison between
the Milky Way, LMC, Ho II and other galaxies will be published elsewhere. 

We conclude that our automatic identification code is a powerful tool in 
studies of \hi shells, which enabled us to perform the so far most extensive 
statistical study of shells in the Milky Way.

\begin{acknowledgements}
The authors gratefully acknowledge financial support by the Grant Agency of 
the Academy of Sciences of the Czech Republic under the grant No. B3003106 
and support by the grant project of the Academy of Sciences of the Czech 
Republic No. K1048102. They would also like to express their thanks to
Bruce G. Elmegreen for comments on how to improve the English and to the 
anonymous referee for suggestions on how to improve the paper.
\end{acknowledgements}

{}

\begin{appendix}

\section{A comparison with  other lists}

\subsection{Heiles}

\begin{figure*}
 \centering
 \includegraphics[angle=0,width=7.5cm]{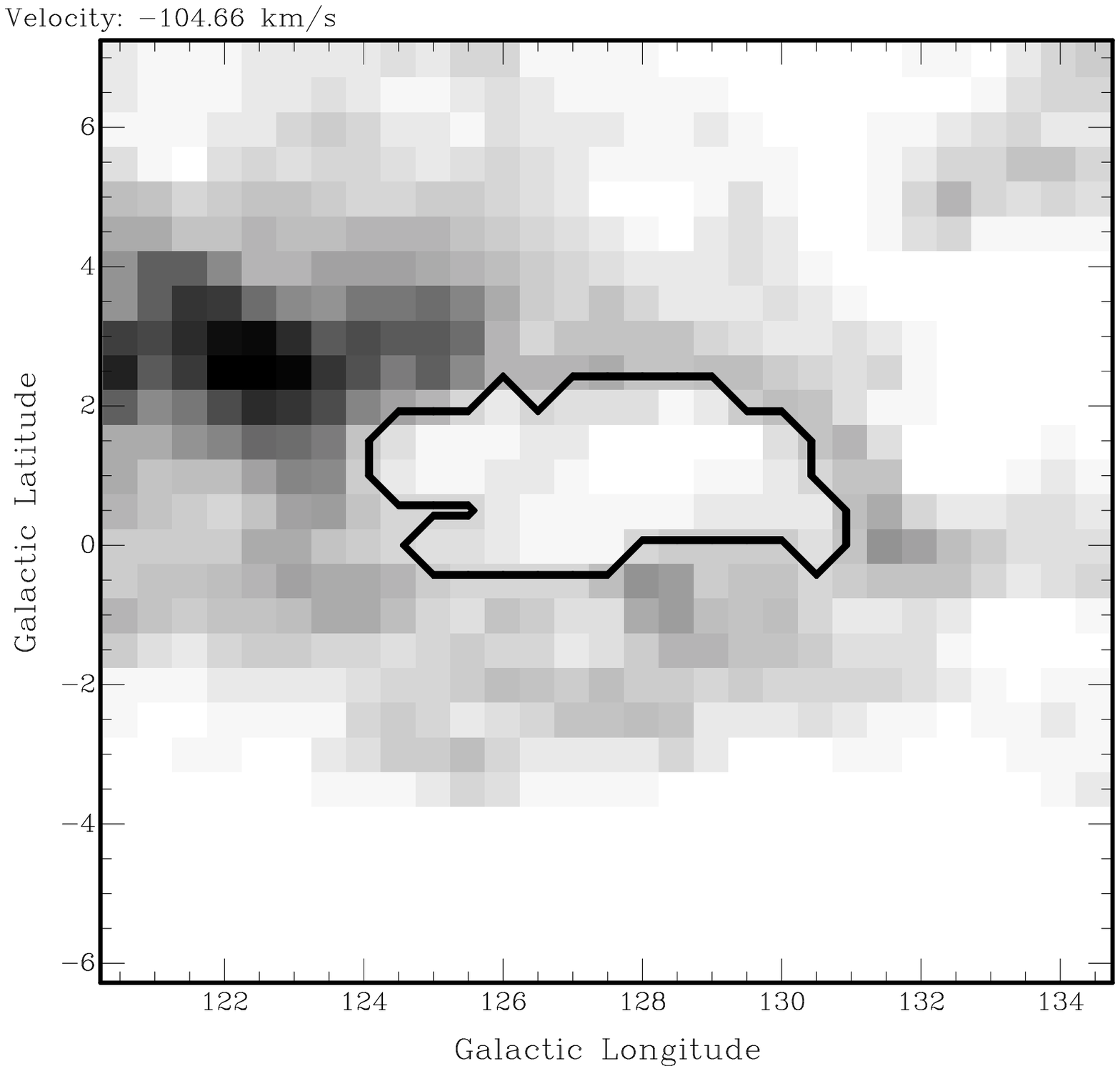}
 \includegraphics[angle=0,width=7.5cm]{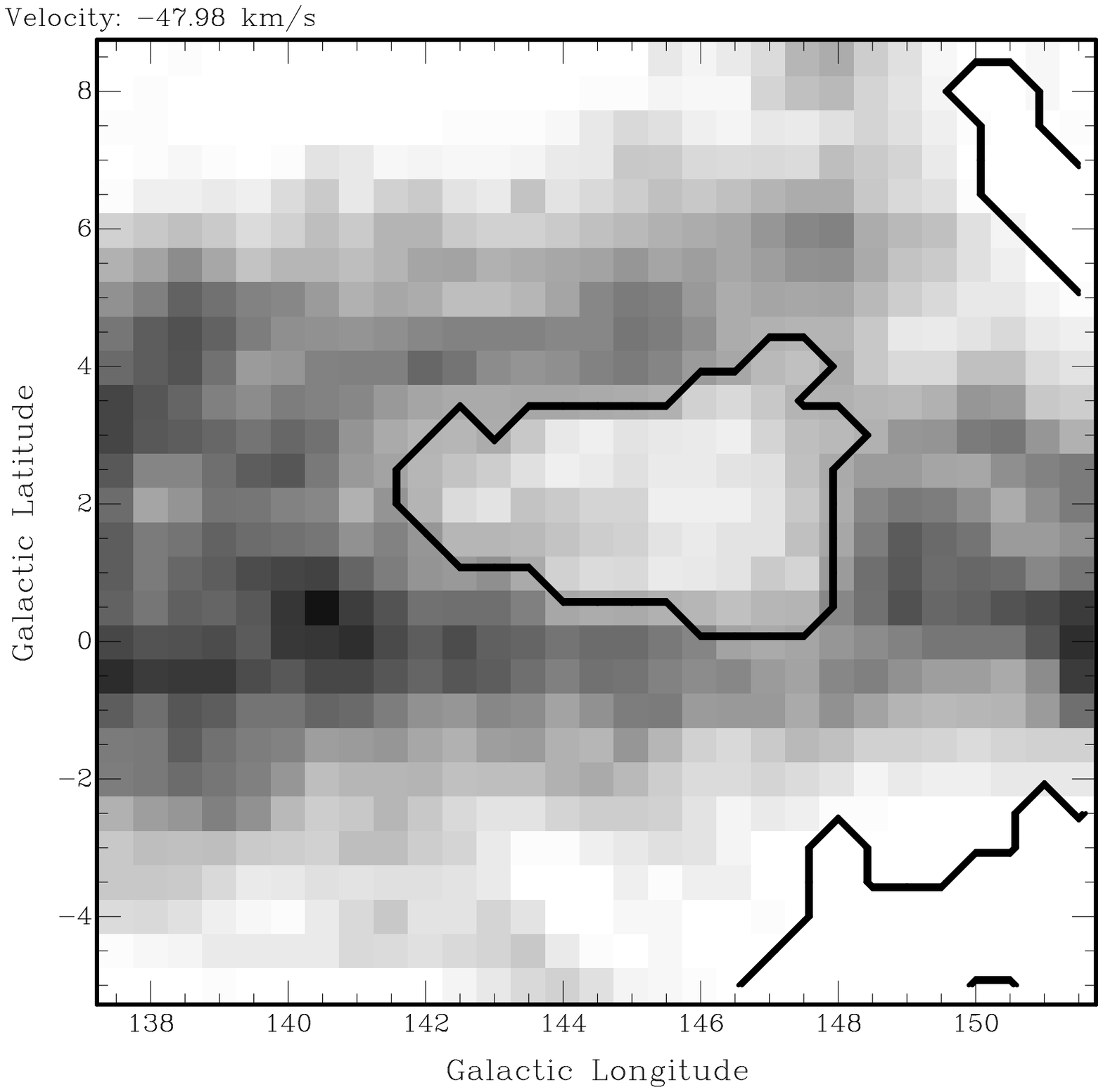}

 \includegraphics[angle=0,width=7.5cm]{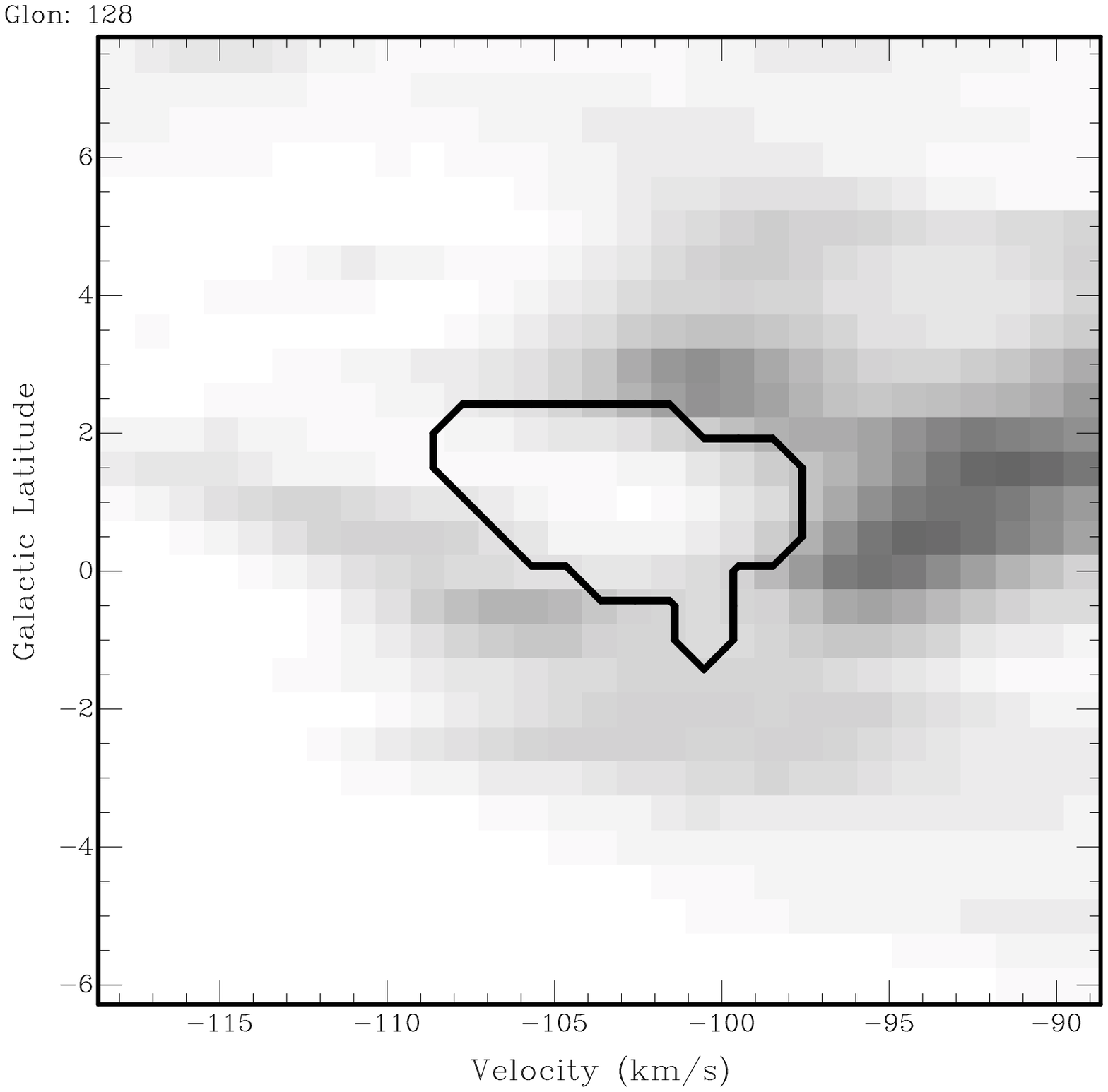}
 \includegraphics[angle=0,width=7.5cm]{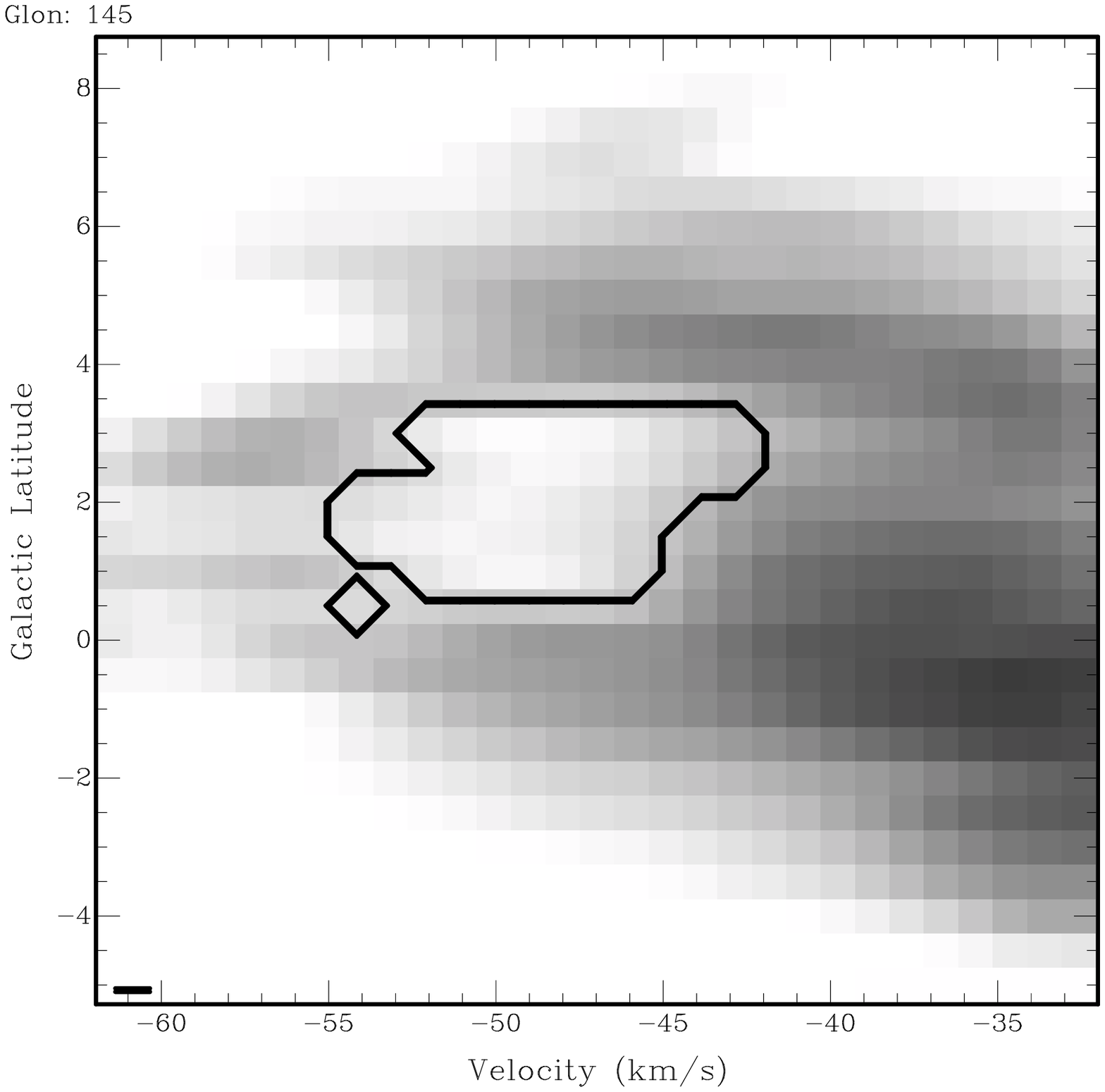}
 \caption{Re-identification of the shell GS 128+01-105 discovered by
          Heiles (1979) - left column, and a newly discovered shell 
          GS145+02-048 (Table 1) - right column. The upper row shows
          $lb-$plots, the bottom row shows $bv-$diagrams.
          }
 \label{figyn}
\end{figure*}

Heiles' list of \hi shells in the Milky Way (Heiles 1979) contains 
fourteen shells in the 2nd Galactic quadrant identified in the \hi 
survey of Weaver \& Williams (1973). A visual inspection of the LDS shows 
that nine of the fourteen shells in the Heiles list are clearly and easily 
seen. Three shells are formed by partial walls and there are no well--defined 
holes. Two shells are completely open to the halo.  

Seven of the nine Heiles structures were re-identified by our method. 
The remaining two were found in velocity channels but were later rejected 
because of unsuitable shapes of spectra. The derived properties of these
nine structures (seven shells and two rejected candidates) are summarized
in Table \ref{tabheilcomp} together with published values. The table shows a 
good correspondence both in angular dimensions and velocity extent. There are 
two cases where previous and newly found dimensions differ: GS091-04-69 and 
GS128+01-105. GS091-04-69 is a shell which has inner walls --- it might 
consist of several smaller shells --- and we do not identify all of them as
parts of the structure. The other shell, GS128+01-105, is found to have a 
larger dimension in $l$ and a larger range in radial velocity than 
was previously thought (see Fig. \ref{figyn} -- left column and also 
Table \ref{tabheilcomp}). Judging by eye we think that in this case 
our derived dimensions are better.

Three shells with partial walls were not identified as shells with our 
automatic method. We doubt the existence of one of them, 
GS123+07-127, as there is only a scarce group of fragments at the indicated 
position. The second shell, GS103+05-137, is very fragmented and we do not 
have any identification at this position, but we find a shell at these $l,b$ 
coordinates in an adjacent velocity interval. According to Heiles, this shell 
is seen at velocities lower than $-123\ \mathrm{kms^{-1}}$ and we see a 
structure in an interval of $(-123,-100)\ \mathrm{kms^{-1}}$. Very probably 
both identifications, Heiles' and ours, are part of the same structure. The 
third structure from the list, GS130+00+15 is a large ($35^{\circ}$ in $l$) 
shell formed by a clearly observed wall, however this wall does not 
encircle the whole structure. The hole (the region of minimum 
in the temperature distribution) is not pronounced in this case.

The two open structures (GS117-07-67 and GS129-05-91) are formed by incomplete
arcs and therefore cannot be found by our algorithm.

\begin{table}[bth]
\centering
\begin{tabular}{|lrr|}
\hline
name & Heiles & this paper \\
~ & $\mathrm{deg}^2 \ \mathrm{kms^{-1}}$ 
  & $\mathrm{deg}^2 \ \mathrm{kms^{-1}}$  \\
\hline
GS090+02-115 & $4\times 4\times 16$ &
               $3\times 4\times 12$ \\
GS091-04-69 & $9\times 10\times 28$ &
              $3\times 5\times 33$ \\
GS091+02-101 & $4\times 3\times 12$ &
               $4\times 4\times 33$ \\
GS108-04-23 & $5\times 11\times 24$ &
              $7\times 14\times 20$ \\
GS128+01-105 & $7\times 6\times 4$ &
               $11\times 5\times 16$ \\
GS148-01+15 & $4\times 4\times 4$ &
              $3\times 5\times 10$ \\
GS152-04-41 & $4\times 4\times 12$ &
              $11\times 9\times 12$ \\
\hline
GS095+04-113 & $10\times 5\times 20$ &
               $10\times 6\times 21$ \\
GS139-03-69 & $18\times 10\times 28$ &
              $13\times 6\times 22$ \\
\hline
\end{tabular}
\caption{A comparison of previously known shells (Heiles 1979) 
         and our identifications. The first column gives the names
         of the structures, the second one gives dimensions as
         published by Heiles, the third column gives our dimensions
         ($\Delta l$, $\Delta b$, $\Delta v$).
         }
\label{tabheilcomp}
\end{table}

\subsection{Hu}

Hu (1981) made a list of high latitude shells detected in the \hi survey
of Heiles and Habing (1974). The high latitude shells in Hu's list are more 
difficult to see in LDS data; about half of them 
can barely be distinguished by eye or are simply not seen
at all (this is connected to the method used for Hu's 
identifications). The Hu list contains twelve shells in the second quadrant. 
Our automatic procedure is able to 
recover 
50 \% of them. Some 
shells from Hu's list are too small (No. 26a) or fragmented (No. 26). 
Others (No. 25 and 28) are not connected to any well--pronounced HI 
depressions.

\subsection{CGPS}

Our identification algorithm was  tested on two fields of the CGPS, where
two structures were found in the automatic search by Mashchenko \& St.-Louis 
(2002). In the two data subsets, which were smoothed to a lower angular
resolution, we were able to recover structures Sh2-203 and Sh2-187
with almost identical positions, dimensions and expansion velocities.

\subsection{Summary}

We identified with an automatic algorithm searching in the 2nd galactic 
quadrant in LDS 50\% of the structures identified as \hi shells in a search 
in the Weaver and Williams (1974) survey performed by eye by Heiles 
(1979), and 50\% of the high latitude shells discovered with a filtering 
method by Hu (1981) in the HI survey of Heiles and Habing (1974). 
We did not find shells formed only by incomplete arcs, or not connected to 
a minimum in $lb$ maps. We found many more (nearly 300) shells than those 
contained in the Heiles and Hu's lists. Many of these newly identified shells 
are well defined (i.e. with a regular shape and good contrast against the 
background) and have properties (dimensions, expansion velocities) comparable 
to those previously found. An example of such a structure at the position: 
$(l_{\mathrm{c}}, b_{\mathrm{c}}, v_{\mathrm{LSR}})$ = (145.4, 2.1, -48.3) 
and $(\Delta l, \Delta b, \Delta v)$ = (6.0, 4.5, 10.3) is shown in 
Fig. \ref{figyn} (right column). Thus, it is surprising that they have not 
been previously discovered.

\end{appendix}

\end{document}